\documentclass[longauth]{aa}

\usepackage{graphicx}
\usepackage{adjustbox}
\usepackage{txfonts}

\usepackage[colorlinks=true,citecolor=blue]{hyperref}
\usepackage{orcidlink}

\usepackage{amsmath}
\usepackage{color} 
\usepackage{xcolor} 
\usepackage{makecell} 
\usepackage{subfigure}
\renewcommand{\arraystretch}{1.3} 
\usepackage[para]{threeparttable} 

\newcommand*\pyorbit{\texttt{PyORBIT}}
\newcommand{\thirstee}{\texttt{THIRSTEE}}

\newcommand*\tess{{\it TESS}}

\newcommand{\snr} {\mbox{S/N}}

\newcommand{\prot}{\mbox{P$_{\rm rot}$}}

\newcommand{\teff}{$T_{{\rm eff}}$}
\newcommand{\kms}{\mbox{km\,s$^{-1}$}}
\newcommand{\ms}{\mbox{m s$^{-1}$}}
\newcommand{\gcm}{\mbox{g cm$^{-3}$}}

\newcommand{\logg} {\mbox{log\,{\it g}}}

\newcommand{\mplanet}{\mbox{$M_{\rm p}$}}
\newcommand{\rplanet}{\mbox{$R_{\rm p}$}}
\newcommand{\rhoplanet}{\mbox{$\rho_{\rm p}$}}

\newcommand{\teq}{$T_{{\rm eq}}$}

\newcommand{\mearth}{\mbox{M$_\oplus$}}
\newcommand{\rearth}{\mbox{R$_\oplus$}}

\newcommand{\msun}{\mbox{M$_\odot$}}
\newcommand{\rsun}{\mbox{R$_\odot$}}
\newcommand{\mstar}{\mbox{$M_\star$}}
\newcommand{\rstar}{\mbox{$R_\star$}}

\newcommand{\rhostar}{\mbox{$\rho_\star$}}
\newcommand{\rhosun}{\mbox{$\rho_\odot$}}

\newcommand{\logRHK}{\mbox{$\log {\rm R}^{\prime}_{\rm HK}$}}

\newcommand{\halpha}{\mbox{H$\alpha$}}

\begin{document}

   \title{Sibling sub-Neptunes around sibling M dwarfs: TOI-521 and TOI-912}

   \author{G. Lacedelli\inst{\ref{iac}}\,$^{\href{https://orcid.org/0000-0002-4197-7374}{\protect\includegraphics[height=0.19cm]{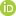}}}$,
          E. Pallé\inst{\ref{iac}, \ref{ull}}\,$^{\href{https://orcid.org/0000-0003-0987-1593}{\protect\includegraphics[height=0.19cm]{orcid.jpg}}}$,
          R. Luque\inst{\ref{iaa}, \ref{chicago}\,\thanks{NHFP Sagan Fellow}}\,$^{\href{https://orcid.org/0000-0002-4671-2957}{\protect\includegraphics[height=0.19cm]{orcid.jpg}}}$,
          K. Ikuta\inst{\ref{hit}}\,$^{\href{https://orcid.org/0000-0002-5978-057X}{\protect\includegraphics[height=0.19cm]{orcid.jpg}}}$,
          H. M. Tabernero\inst{\ref{ieec},\ref{ice_csic}},
          M. R. Zapatero Osorio\inst{\ref{CSIC-INTA}}\,$^{\href{https://orcid.org/0000-0001-5664-2852}{\protect\includegraphics[height=0.19cm]{orcid.jpg}}}$,
          J.M.~Almenara\inst{\ref{Grenoble}}\,$^{\href{https://orcid.org/0000-0003-3208-9815}{\protect\includegraphics[height=0.19cm]{orcid.jpg}}}$,
          F. J. Pozuelos\inst{\ref{iaa}},
          D. Jankowski\inst{\ref{poland}},
          N. Narita\inst{\ref{komabains}, \ref{abc}, \ref{iac}}\,$^{\href{https://orcid.org/0000-0001-8511-2981}{\protect\includegraphics[height=0.19cm]{orcid.jpg}}}$,
          A. Fukui\inst{\ref{komabains}, \ref{iac}}\,$^{\href{https://orcid.org/0000-0002-4909-5763}{\protect\includegraphics[height=0.19cm]{orcid.jpg}}}$,
          G. Nowak\inst{\ref{poland}}\,$^{\href{https://orcid.org/0000-0002-7031-7754}{\protect\includegraphics[height=0.19cm]{orcid.jpg}}}$,
          T. Hirano\inst{\ref{abc}, \ref{naoj},\ref{sokendai}}\,$^{\href{https://orcid.org/0000-0003-3618-7535}{\protect\includegraphics[height=0.19cm]{orcid.jpg}}}$,
          H. T. Ishikawa\inst{\ref{uwo}}\,$^{\href{https://orcid.org/0000-0001-6309-4380}{\protect\includegraphics[height=0.19cm]{orcid.jpg}}}$, 
          T. Kimura\inst{\ref{utops}, \ref{gron}}\,$^{\href{https://orcid.org/0000-0001-8477-2523}{\protect\includegraphics[height=0.19cm]{orcid.jpg}}}$, 
          Y. Hori\inst{\ref{okayama}}\,$^{\href{https://orcid.org/0000-0003-4676-0251}{\protect\includegraphics[height=0.19cm]{orcid.jpg}}}$,
          K. A. \ Collins\inst{\ref{cambridge}}\,$^{\href{https://orcid.org/0000-0001-6588-9574}{\protect\includegraphics[height=0.19cm]{orcid.jpg}}}$,
          S.~B.~Howell\inst{\ref{NASA}}\,$^{\href{https://orcid.org/0000-0002-2532-2853}{\protect\includegraphics[height=0.19cm]{orcid.jpg}}}$,
          C. Jiang\inst{\ref{iac}, \ref{ull}},
          F. Murgas\inst{\ref{iac},\ref{ull}}\,$^{\href{https://orcid.org/0000-0001-9087-1245}{\protect\includegraphics[height=0.19cm]{orcid.jpg}}}$,
          H. P. Osborn\inst{\ref{bern},\ref{ETH}},
          N. Astudillo-Defru \inst{\ref{chile}}\,$^{\href{https://orcid.org/0000-0002-8462-515X}{\protect\includegraphics[height=0.19cm]{orcid.jpg}}}$,
          X. Bonfils \inst{\ref{Grenoble}},
          D. Charbonneau\inst{\ref{cambridge}}\,$^{\href{https://orcid.org/0000-0002-9003-484X}{\protect\includegraphics[height=0.19cm]{orcid.jpg}}}$,
          M.~M.~Fausnaugh \inst{\ref{texas}}\,$^{\href{https://orcid.org/0000-0002-9113-7162}{\protect\includegraphics[height=0.19cm]{orcid.jpg}}}$,
          S.~Gerald\'ia-Gonz\'alez\inst{\ref{iac},\ref{ull}},$^{\href{https://orcid.org/0009-0002-5545-3034}{\protect\includegraphics[height=0.19cm]{orcid.jpg}}}$
          K. Goździewski \inst{\ref{poland}},
          P. Guerra \inst{\ref{girona}},$^{\href{https://orcid.org/0000-0002-0619-7639}{\protect\includegraphics[height=0.19cm]{orcid.jpg}}}$, 
          Y. Hayashi\inst{\ref{komaba}}\,$^{\href{https://orcid.org/0000-0001-8877-0242}{\protect\includegraphics[height=0.19cm]{orcid.jpg}}}$, 
          K. Hodapp\inst{\ref{hawaii}}\,$^{\href{https://orcid.org/0000-0003-0786-2140}{\protect\includegraphics[height=0.19cm]{orcid.jpg}}}$,  
          K. Horne \inst{\ref{scotland}}\,$^{\href{https://orcid.org/0000-0003-1728-0304}{\protect\includegraphics[height=0.19cm]{orcid.jpg}}}$, 
          K. Isogai\inst{\ref{kyoto},\ref{komaba}}\,$^{\href{https://orcid.org/0000-0002-6480-3799}{\protect\includegraphics[height=0.19cm]{orcid.jpg}}}$,  
          M.~Jafariyazani \inst{\ref{SETI}, \ref{NASA}}\,$^{\href{https://orcid.org/0000-0002-9113-7162}{\protect\includegraphics[height=0.19cm]{orcid.jpg}}}$,  
        T. Kagetani\inst{\ref{komaba},\ref{naoj}}\,$^{\href{https://orcid.org/0000-0002-5331-6637}{\protect\includegraphics[height=0.19cm]{orcid.jpg}}}$, 
        Y. Kawai\inst{\ref{komaba}}\,$^{\href{https://orcid.org/0000-0002-0488-6297}{\protect\includegraphics[height=0.19cm]{orcid.jpg}}}$, 
        K. Kawauchi\inst{\ref{ritsumei}}\,$^{\href{https://orcid.org/0000-0003-1205-5108}{\protect\includegraphics[height=0.19cm]{orcid.jpg}}}$, 
        V. Krishnamurthy\inst{\ref{mcgill}}\,$^{\href{https://orcid.org/0000-0003-2310-9415}{\protect\includegraphics[height=0.19cm]{orcid.jpg}}}$, 
        T. Kotani\inst{\ref{abc}, \ref{naoj},\ref{sokendai}}\,$^{\href{https://orcid.org/0000-0001-6181-3142}{\protect\includegraphics[height=0.19cm]{orcid.jpg}}}$,
        T. Kudo\inst{\ref{subaru}}\,$^{\href{https://orcid.org/0000-0002-9294-1793}{\protect\includegraphics[height=0.19cm]{orcid.jpg}}}$, 
        T. Kurokawa\inst{\ref{naoj},\ref{agri}},
        M. Kuzuhara\inst{\ref{abc}, \ref{naoj}}\,$^{\href{https://orcid.org/0000-0002-4677-9182}{\protect\includegraphics[height=0.19cm]{orcid.jpg}}}$, 
        M. Mori\inst{\ref{abc}, \ref{naoj}}\,$^{\href{https://orcid.org/0000-0003-1368-6593}{\protect\includegraphics[height=0.19cm]{orcid.jpg}}}$, 
        J. Nishikawa\inst{\ref{naoj},\ref{sokendai},\ref{abc}}\,$^{\href{https://orcid.org/0000-0001-9326-8134}{\protect\includegraphics[height=0.19cm]{orcid.jpg}}}$, 
        S. K. Nugroho\inst{\ref{abc}, \ref{naoj}}\,$^{\href{https://orcid.org/0000-0003-4698-6285}{\protect\includegraphics[height=0.19cm]{orcid.jpg}}}$,  
        M. Omiya\inst{\ref{abc}, \ref{naoj}}, 
        R. P. Schwarz \inst{\ref{cambridge}}\,$^{\href{https://orcid.org/0000-0001-8227-1020}{\protect\includegraphics[height=0.19cm]{orcid.jpg}}}$, 
        R. Sefako \inst{\ref{south_africa}},
        A.~Shporer \inst{\ref{kavli}}\,$^{\href{https://orcid.org/0000-0002-1836-3120}{\protect\includegraphics[height=0.19cm]{orcid.jpg}}}$, 
        G. Srdoc\inst{\ref{croatia}}, 
        H. Teng\inst{\ref{cas}}\,$^{\href{https://orcid.org/0000-0003-3860-6297}{\protect\includegraphics[height=0.19cm]{orcid.jpg}}}$, 
        N. Watanabe\inst{\ref{komaba}}\,$^{\href{https://orcid.org/0000-0002-7522-8195}{\protect\includegraphics[height=0.19cm]{orcid.jpg}}}$
          }

   \institute{Instituto de Astrof\'{i}sica de Canarias (IAC), 38205 La Laguna, Tenerife, Spain \label{iac} 
              \email{glacedelli@iac.es}
    \and Departamento de Astrof\'isica, Universidad de La Laguna (ULL), E-38206 La Laguna, Tenerife, Spain \label{ull} 
    \and Instituto de Astrof\'isica de Andaluc\'ia (IAA-CSIC), Glorieta de la Astronom\'ia s/n, 18008 Granada, Spain \label{iaa}
    \and Department of Astronomy \& Astrophysics, University of Chicago, Chicago, IL 60637, USA \label{chicago}
    \and Department of Social Data Science, Hitotsubashi University, 2-1 Naka, Kunitachi, Tokyo 186-8601, Japan \label{hit}
    \and Institut d'Estudis Espacials de Catalunya (IEEC), Edifici RDIT, Campus UPC, 08860 Castelldefels (Barcelona), Spain \label{ieec}
    \and Institut de Ciències de l'Espai (ICE, CSIC), Campus UAB, c/ de Can Magrans s/n, 08193 Cerdanyola del Vallès, Barcelona, Spain \label{ice_csic}
    \and Centro de Astrobiología (CSIC-INTA), Crta. Ajalvir km 4, E-28850 Torrejón de Ardoz, Madrid, Spain \label{CSIC-INTA}
    \and Univ. Grenoble Alpes, CNRS, IPAG, F-38000 Grenoble, France\label{Grenoble}
    \and Institute of Astronomy, Faculty of Physics, Astronomy and Informatics, Nicolaus Copernicus University, Grudzi\c{a}dzka 5, 87-100 Toru\'{n}, Poland \label{poland}
    \and Komaba Institute for Science, The University of Tokyo, 3-8-1 Komaba, Meguro, Tokyo 153-8902, Japan \label{komabains}
    \and Astrobiology Center, 2-21-1 Osawa, Mitaka, Tokyo 181-8588, Japan \label{abc}
    \and National Astronomical Observatory of Japan, 2-21-1 Osawa, Mitaka, Tokyo 181-8588, Japan \label{naoj}
    \and Department of Astronomical Science, The Graduate University for Advanced Studies (SOKENDAI), 2-21-1 Osawa, Mitaka, Tokyo 181-8588, Japan \label{sokendai}
    \and Department of Physics and Astronomy, The University of Western Ontario, 1151 Richmond St, London, Ontario, N6A 3K7, Canada \label{uwo}
    \and UTokyo Organization for Planetary and Space Science,The University of Tokyo, 7-3-1 Hongo, Bunkyo, Tokyo 113-0033, Japan \label{utops}
    \and Kapteyn Astronomical Institute, University of Groningen, P.O. Box 800, 9700 AV Groningen, Netherlands \label{gron}
    \and Graduate School of Environmental, Life, Natural Science and Technology, Okayama University, 3-1-1 Tsushima-naka, Kita, Okayama 700-8530, Japan \label{okayama}
    \and Center for Astrophysics \textbar \ Harvard \& Smithsonian, 60 Garden Street, Cambridge, MA 02138, USA \label{cambridge}
    \and NASA Ames Research Center, Moffett Field, CA 94035, USA \label{NASA}
    \and Centre for Space and Habitability, Universität Bern, Gesellschaftsstrasse 6, 3012 Bern, Switzerland \label{bern}
    \and ETH Zurich, Department of Physics, Wolfgang-Pauli-Strasse 2, CH-8093 Zurich, Switzerland \label{ETH}
    \and Departamento de Matem\'atica y F\'isica Aplicadas, Universidad Cat\'olica de la Sant\'isima Concepci\'on, Alonso de Rivera 2850, Concepci\'on, Chile \label{chile}
    \and Department of Physics \& Astronomy, Texas Tech University, Lubbock TX, 79409-1051, USA \label{texas}
    \and Observatori Astronòmic Albanyà, Camí de Bassegoda S/N, Albanyà 17733, Girona, Spain \label{girona}
    \and Department of Multi-Disciplinary Sciences, The University of Tokyo, 3-8-1 Komaba, Meguro, Tokyo 153-8902, Japan \label{komaba}
    \and Institute for Astronomy, University of Hawaii, 640 N. Aohoku Place, Hilo, HI 96720, USA \label{hawaii}
    \and SUPA Physics and Astronomy, University of St. Andrews, Fife, KY16 9SS Scotland, UK \label{scotland}
    \and Okayama Observatory, Kyoto University, 3037-5 Honjo, Kamogata, Asakuchi, Okayama 719-0232, Japan \label{kyoto}
    \and SETI Institute, Mountain View, CA 94043 USA \label{SETI}
    \and Department of Physical Sciences, Ritsumeikan University, 1-1-1 Noji-higashi, Kusatsu, Shiga 525-8577, Japan \label{ritsumei}
    \and Trottier Space Institute at McGill, McGill University, 3550 University Street, Montreal, QC H3A 2A7, Canada \label{mcgill}
    \and Subaru Telescope, National Astronomical Observatory of Japan, 650 N. Aohoku Place, Hilo, HI 96720, USA \label{subaru}
    \and Institute of Engineering, Tokyo University of Agriculture and Technology, 2-24-26 Nakacho, Koganei, Tokyo, 184-8588, Japan \label{agri}
    \and South African Astronomical Observatory, P.O. Box 9, Observatory, Cape Town 7935, South Africa \label{south_africa}
    \and Department of Physics and Kavli Institute for Astrophysics and Space Research, Massachusetts Institute of Technology, Cambridge, MA 02139, USA \label{kavli}
    \and Kotizarovci Observatory, Sarsoni 90, 51216 Viskovo, Croatia \label{croatia}
    \and CAS Key Laboratory of Optical Astronomy, National Astronomical Observatories, Chinese Academy of Sciences, Beijing 100101, China \label{cas}
             }

   \date{Received 18 September 2025; accepted 21 November 2025}

 
  \abstract
   {Sub-Neptunes are absent in the Solar System, yet they are the most common category of planets found in our Galaxy. This kind of planet challenges the internal structure models and prompts investigation on their formation and evolution, as well as pushing towards atmospheric characterisation studies to break the degeneracy in their inner composition. }
   {We report here the discovery and characterisation of new sub-Neptunes orbiting two similar M dwarfs, TOI-521 (\teff~$=3544 \pm 100$~K, $V = 14.7$~mag), and TOI-912 (\teff~$=3572 \pm 100$~K, $V = 12.7$~mag). Both stars host a transiting planetary candidate identified by \textit{TESS} and are part of the \thirstee\ follow-up program, which aims at understanding the sub-Neptune population through in-depth and precise characterisation studies on a population level.}
   {We analysed \tess\ light curves, ground-based photometry and high-precision ESPRESSO, HARPS and IRD radial velocities to confirm the planetary nature of both candidates, infer precise orbital and physical parameters of the planets and investigate the presence of additional planets in the systems.}
   {The two stars host nearly identical planets in terms of mass and radius. TOI-521 hosts a transiting sub-Neptune in a $1.5$-days orbit with radius and mass of $R_{\rm b} = 1.98 \pm 0.14 $~\rearth\ and $M_{\rm b} = 5.3 \pm 1.0$~\mearth, respectively. Moreover, we identified an additional candidate at $20.3$ days, with a minimum mass of $M_{\rm p} \sin{i} = 10.7_{-2.4}^{+2.5}$~\mearth, currently not detected to transit in our photometric dataset. Similarly, the planet orbiting TOI-912 is a $4.7$-d sub-Neptune with $R_{\rm b} = 1.93 \pm 0.13 $~\rearth\ and $M_{\rm b} = 5.1 \pm 0.5$~\mearth. Interestingly, TOI-912 b likely possesses an unusually high eccentricity ($e = 0.58 \pm 0.02$), and it is probably undergoing strong tidal dissipation. If such eccentricity is confirmed, it would make it one of the most eccentric sub-Neptunes known to date. TOI-521 b and TOI-912 b have very similar densities ($\sim 4$~\gcm), and they lie in the degenerate region of the mass-radius diagram where different compositions are plausible, including a volatile-rich composition, or a rocky core surrounded by a H-He envelope. When compared to the other \thirstee\ M-dwarf targets, our sample supports the division of sub-Neptunes into two distinct populations divided by a density gap. Both planets are interesting targets for atmospheric follow-up in the context of understanding the temperature-atmospheric feature trend that starts to emerge thanks to \textit{JWST} observations. 
   }
  {}

\titlerunning{TOI-521 \& TOI-912: two new well-characterised sub-Neptunes}
\authorrunning{Lacedelli et al.}
   \maketitle
%
\section{Introduction}\label{sec:intro}
The origin and nature of the sub-Neptune population is currently one of the big open questions in the exoplanetary community. These small planets (1 \rearth < \rplanet < 4 \rearth) are ubiquitous in the solar vicinity (e.g. \citealt{Batalha2013, Petigura13PNAS, biazzo2022}), yet no such a planet exists in our Solar System. 
When considering their radius distribution, smaller planets, the so-called super-Earths (\rplanet $\lesssim 1.4$~\rearth) are commonly thought to have a rocky composition, while planets with bigger radii, especially in the $1.4$~\rearth\ $\lesssim$ \rplanet\ $\lesssim$ 2 \rearth\ regime, could be explained assuming different internal compositions, and different formation and evolution processes \citep{RogersSeager10}. 
The two most common theories to explain the distribution and properties of these planets are: 1) based on atmospheric mass-loss processes (e.g. \citealt{OwenWu13, ginzburg2018, gupta2019, Rogers23}), according to which these planets are rocky cores surrounded by a variable amount of H/He atmospheres, or 2) 
based on volatile-rich theories (e.g. \citealt{Bitsch19, Izidoro22, burn2021}) which claim that these planets formed outside the water ice line, and they are therefore rich in water, in quantities up to $50$\% of their total mass. 
Currently, detailed models assuming more realistic and complex conditions for the planetary interiors are being developed, including
atmosphere-magma ocean interactions \citep{Schlichting_2022, misener23, Rogers2024, gupta2025}, effects of high irradiation on atmospheres and water content \citep{turbet2020, aguichine2021}, and miscibility of water in the planetary interior \citep{DornLichtenberg21, Luo24}.
In this context, atmospheric study on well-characterised planets are of primary importance, as the study of atmospheric features can provide additional observational constraint to unveil the composition and history of sub-Neptunes.
The initial \textit{JWST} studies of sub-Neptunes atmospheres have already shown promising results, highlighting the detection of atmospheric features and tentative trends within this population \citep{Madhusudhan_2023, bennecke2024, Piaulet2024, davenport2025}.

Within this active and growing field, the \thirstee\ project, described in \cite{lacedelli2024}, is currently trying to shed light on the nature and origin of sub-Neptunes through an observational approach, enlarging the sample of well-characterised sub-Neptunes orbiting both M-dwarf and FGK stars and studying their atmospheres, with the final aim of performing demographic analysis on such systems taking into account all their observed properties. Here, we present the discovery and characterization of two new sub-Neptunes orbiting the M-dwarf stars TOI-521 and TOI-912, as part of the \thirstee\ sample.

\section{TESS photometry}\label{sec:tess}

{\it TESS} observed TOI-521 and TOI-921 for a total of seven and six sectors, respectively (see summary of observations in Table~\ref{table:photometric_observations}). 
The Science Processing Operations Center (SPOC) pipeline at NASA Ames Research Center \citep{jenkins2016,jenkins2020} reduced and analysed image data and identified a transiting candidate around each star, namely TOI-521.01 (with period $P \simeq 1.54$ d) and TOI-912.01 ($P \simeq 4.68$ d).

For both stars we analysed the two-minute cadence Pre-search Data Conditioning Simple Aperture Photometry
\citep[PDCSAP, ][]{smith2012, Stumpe2012, Stumpe2014} light curves, reduced with the SPOC pipeline (see field of view of the targets and photometric aperture in Fig.~\ref{fig:tpf}).
For the photometric analysis, we selected from the PDCSAP only `good quality' data points (\texttt{DQUALITY=0}\footnote{ \url{https://archive.stsci.edu/missions/tess/doc/EXP-TESS-ARC-ICD-TM-0014.pdf}}), 
and we clipped $3\sigma$ outliers, masking the in-transit points according to the ephemeris provided on the \tess\ Follow-up Observing Program (TFOP\footnote{\url{https://tess.mit.edu/followup}}; \citealt{collins2019}) to avoid affecting the transits shape.

\section{Ground-based follow-up observations}\label{sec:ground_based}
We report here all the ground-based observations used in this work, including the ground-based transit observations collected as part of TFOP, long-term photometric monitoring, and the high-precision spectroscopic observations. Observations are summarised in Table~\ref{table:photometric_observations}, and all ground-based transit light-curves are shown in Fig.~\ref{fig:ground_based}.
Appendix~\ref{appendix:light_curves} also reports the high-resolution speckle imaging we carried out to provide confirmation for TOI-521.01 and TOI-912.01, which reveals no close companions to both stars within the angular and magnitude limits.

\subsection{Transit photometry}\label{sec:extra}

\paragraph{ExTrA} We observed TOI-912.01 with the Exoplanets in Transits and their Atmospheres (ExTrA) facility \citep{bonfil2015} located at La Silla Observatory, Chile. We collected ten transits during separate nights, but various transits were simultaneously observed by more than one of the three $0.6$-m telescopes in the facility, resulting in a total of 23 light curves. 
All observations were gathered with the $8$-arcsec fibre, in the low-resolution mode of the spectrograph, and with an exposure time of $60$ seconds. Data were reduced with the custom reduction software detailed in \cite{cointepas2021}.

\paragraph{LCOGT} We observed three full transit windows of TOI-521.01 from the Las Cumbres Observatory Global Telescope (LCOGT) \citep{brown2013} 1\,m network nodes at McDonald Observatory near Fort Davis, Texas, United States (McD), Siding Spring Observatory near Coonabarabran, Australia (SSO), and Cerro Tololo Inter-American Observatory in Chile (CTIO). We also observed three full transits of TOI-912.01 from LCOGT 1\,m network nodes at South Africa Astronomical Observatory near Sutherland, South Africa (SAAO), CTIO, and SAAO, respectively. 
The images were calibrated using the standard LCOGT {\tt BANZAI} pipeline \citep{McCully2018} and differential photometric data were extracted using {\tt AstroImageJ} \citep{collins2017}. We used circular photometric apertures with radii in the range $5\farcs0$ to $7\farcs3$, which excluded flux from all known \textit{Gaia} DR3 catalogue neighbours that are bright enough to have hosted a blended nearby eclipsing binary that could have masqueraded as the \textit{TESS} detected transit.

\paragraph{MuSCAT2} We collected one full transit of TOI-521.01 with the multi-colour camera MuSCAT2 \citep{narita2019}, mounted on the Telescopio Carlos Sànchez at the Teide Observatory. The exposure time of the four MuSCAT2 bands ($g$-, $r$-, $i$-, and $z_s$) was set to 15, 30, 30, and 45 s, respectively. 
Data were reduced and calibrated with the MuSCAT2 pipeline \citep{Parviainen2019}, which optimizes the photometric aperture while performing the transit fit.

\paragraph{MEarth} We observed a full transit of TOI-912.01 with the MEarth South facility \citep{nutzman2008, irwin2015} at the Cerro Tololo Inter-American Observatory, Chile. 
We extracted the photometry with a circular aperture of $7\farcs 1$ through a custom pipeline outlined in \cite{irwin2007} and \cite{berta2012}.

\subsection{Long-term photometric monitoring}\label{sec:asas}
We used the All-Sky Automated Survey for Supernovae (ASAS-SN\footnote{\url{http://asas-sn.ifa.hawaii.edu/skypatrol/} \citep{hart2023}} \citep{Shappee2014, Kochanek2017} and Zwicky Transient Facility\footnote{\url{https://irsa.ipac.caltech.edu/Missions/ztf.html}} (ZTF, \citealt{Bellm:2019,Masci:2019}) long-term photometry to investigate stellar activity and search for the rotational period of the stars. 
TOI-912 ASAS-SN observations span between 2014-2018 ({\it V} filter), and 2018-2024 ({\it g} filter), for a total time span of $\sim 1583$ and $\sim 2318$ days, respectively. No ZTK observations are available for this star.

TOI-521 was observed by ASAS-SN between 2012-2018 ({\it V} filter), and between 2017-2024 ({\it g} filter) for a total time span of $\sim 2505$ and $\sim 2423$ days, respectively.

The star was observed in the ZTF $r$ filter between 2018 and 2024, for a total baseline of $\sim 2216$~d. 
For all light curves, we selected only observations flagged as good-quality (\texttt{QUALITY=G} for ASAS-SN, and \texttt{catflags=0} for ZTK), and we performed a $5 \sigma$ clipping to remove outliers. 
Figs.~\ref{fig:asas_sn_toi521} and \ref{fig:asas_sn_TOI912} show the resulting light curves.

\subsection{Spectroscopic observations}\label{sec:spectra}

\paragraph{ESPRESSO} We followed-up both stars within the \thirstee\ program with the ESPRESSO  spectrograph \citep{pepe2021} on the ESO Very Large Telescope telescope array. 
We used ESPRESSO's high-resolution mode ($1$~arcsec fibre, $R \sim 140 000$) with the $2 \times 1$ detector binning (HR21) on a single Unit Telescope (1UT), and we placed fibre B on the sky for background subtraction.

For TOI-521, we collected $22$ data points (Program 112.25F2.001, PI: E. Palle) between November 13, 2023, and March 15, 2024, spanning a total of $\sim 123$ days.
The adopted exposure time was $1800$ seconds, implying a median signal-to-noise  ratio (\snr) of $19$ at $550$~nm. We reduced the data with the ESPRESSO's data reduction pipeline (DRS), and we extracted the RVs using a M3 mask to compute the cross-correlation function (CCF). 
In our global analysis, we removed $1$ spectrum which was collected in bad observing conditions, resulting in low \snr\ ratio and with an associated internal uncertainty $> 3$~\ms\ ($3\sigma$ outlier). 

For TOI-912, we collected $33$ data points (Programs 111.24PJ.001 and 113.26NP.001, PI: E. Palle), between April 10, 2023, and August 11, 2024, spanning a total of $\sim 488$ days.
The exposure time was $900$ seconds, implying a median \snr\ of $31$ at $550$~nm. Given the higher number of spectra, and the brightness of the target, which allows for a high \snr\ spectral template, in this case we extracted the RVs with the template-matching \texttt{SERVAL}\footnote{\url{https://github.com/mzechmeister/serval}.} algorithm \citep{zechmeister2018}, which provided more precise median internal uncertainties and smaller RMS.  

The resulting RV datasets for the two stars (available on CDS) have an RMS of $5.3$~\ms\ and $2.8$~\ms\, and a median internal uncertainty of $1.5$~\ms and $0.47$~\ms, respectively.

\paragraph{IRD} We collected 65 RVs for TOI-521 with the InfraRed Doppler (IRD) spectrograph at the Subaru 8.2-m telescope \citep{tamura2012, kotani2018}, under the Subaru/IRD intensive programs S19A-069I, S20B-088I, S21B-118I, and S23A-067I (PI: N. Narita).
The observing campaign spanned between April 18, 2019 and April 7, 2023, for a total observational baseline of $\sim 1450$~days. 
The integration time per exposure ranged from 180 to 1800 s, depending on the observing condition for each night, and the raw data were reduced following \cite{Kuzuhara18, Hirano20}, while RVs were extracted as outlined in \citet{Hirano20,Kuzuhara24}. 
We excluded from the dataset three points collected in bad observing conditions, resulting in low \snr\ ratio. 
The final RV dataset (available on CDS) has an RMS of $14.4$~\ms, and a median internal uncertainty of $7.6$~\ms.
We also computed several indicators to investigate the stellar activity, namely the full width at half maximum (FWHM), the BiGauss (dV; the line asymmetry) by fitting Gaussian functions \citep{Santerne15}, the chromatic index (CRX; the wavelength dependence of RV) and the differential line width (dLW, \citealt{zechmeister2018}) as in \cite{harakawa2022}.

\paragraph{HARPS}
We collected 22 RVs for TOI-912 (program 1102.C-0339(A), PI: Bonfils) between July 28, 2019 and September 14, 2019 ($\sim 49$~d baseline) with the HARPS \citep{Pepe_2002}, on the ESO 3.6-m telescope at La Silla. 
We adopted the HARPS configuration with the second fibre on the sky, and an exposure time of $1800$ seconds. Data were reduced with the HARPS DRS (version 3.8), and we extracted the RVs with the \texttt{NAIRA} pipeline \citep{astudillo_defru_2017_naira}. 
The resulting RV dataset (available on CDS) has an RMS of $4.2$~\ms, and a median internal uncertainty of $3.1$~\ms.

\section{The stars}\label{sec:star}

\begin{table*}
\small
\centering
\caption{Stellar properties of TOI-521 and TOI-912.}
\label{table:star_params}
\begin{threeparttable}[t]
\begin{tabular}{lllc}
\hline\hline
Parameter & TOI-521 & TOI-912 &  Source\\
\hline
TIC & 27649847 & 406941612 &  \\
{\it Gaia} DR3 & 649852779797683968 & 5772442647192375808\\
2MASS & J08132251+1213181 &  J15172165-8028225 \\
\hline
RA  (J2000; hh:mm:ss.ss) &  08:13:22.62 & 15:17:18.76 & A\\
Dec (J2000; dd:mm:ss.ss) & $+$12:13:19.63 & $-$80:28:24.09& A\\
$\mu_{\alpha}$ (mas yr$^{-1}$) & $90.415 \pm 0.019 $ & $-451.119 \pm 0.016 $ & A\\
$\mu_{\delta}$ (mas yr$^{-1}$) & $80.780 \pm 0.013$ & $-93.820 \pm 0.017 $ & A\\
Parallax (mas) & $16.388 \pm 0.020$& $38.327 \pm 0.015$ & A\\
Distance (pc) & $60.84_{-0.08}^{+0.07}$& $26.089_{-0.009}^{+0.008}$ &B \\
$\gamma$ (\kms) & $-9.31 \pm 1.10$ & $10.47 \pm 0.31$ & A \\
U (\kms) & $13.47 \pm 0.86$& $-30.29 \pm 0.18$& F$^a$  \\ 
V (\kms) & $20.17 \pm 0.52$& $ -46.38  \pm 0.23$& F$^a$  \\
W (\kms) & $26.85 \pm 0.44$& $16.98 \pm 0.10$& F$^a$  \\
\hline
{\it TESS} (mag) & $12.403 \pm 0.007$& $10.490 \pm 0.007$ & C\\
{\it G} (mag) & $13.568 \pm 0.003$ & $ 11.643 \pm 0.003$ & A\\
{\it G$_{\rm BP}$} (mag) & $14.878 \pm 0.004$ & $12.928 \pm 0.003$ & A\\
{\it G$_{\rm RP}$} (mag) & $12.436\pm 0.004$ & $10.520 \pm 0.004$ & A\\
{\it B } (mag) & $16.268 \pm 0.147$ & $14.170 \pm 0.018$ & C\\
{\it V }(mag) & $14.691 \pm 0.066$ & $12.670 \pm 0.048$ &C\\
{\it J }(mag) & $10.921 \pm 0.022$ & $9.089 \pm 0.023$ & D\\
{\it H }(mag) & $10.357 \pm 0.020$ & $8.468 \pm 0.040$ & D\\
{\it K }(mag) & $10.089 \pm 0.0191$ & $8.204\pm 0.023$ & D\\
{\it W1} (mag) & $9.985 \pm 0.023$ & $8.033 \pm 0.024$ & E\\
{\it W2} (mag) & $9.854 \pm 0.021$ & $7.943 \pm 0.021$ & E\\
{\it W3} (mag) & $9.727 \pm 0.0474$ & $7.831 \pm 0.018$ & E\\
{\it W4} (mag) & $ >9.006 $ & $7.630 \pm 0.113$ & E\\[1ex]
\hline
\teff\ (K) & $ 3544 \pm 100 $ & $ 3572 \pm 100 $ & F \\ 
\logg\  (cgs)  & $4.97 \pm 0.10$ & $4.98 \pm 0.09$ & F, {\tiny Spectroscopic} \\
\logg\  (cgs)  & $ 4.81 \pm 0.09$ & $  4.82 \pm 0.09$ & F, {\tiny Bolometric} \\ 
$[$Fe/H$]$ (dex) & $-0.12 \pm 0.08$ & $-0.19 \pm 0.07$ &F \\
\logRHK & $  -5.500 \pm  0.036 $ & $  -5.176 \pm 0.011  $ & F\\
$P_{\rm rot}$ (d) & - &$45.9_{-0.2}^{+0.6}$ & F\\
\rstar\ (\rsun) & $ 0.421 \pm 0.028 $ & $ 0.419 \pm 0.028$  & F\\
\mstar\ (\msun) & $ 0.420 \pm 0.030$ & $ 0.418 \pm 0.030$  &F \\
$L_{\star}$ ($L_{\odot}$) & $ 0.0252 \pm  0.0005 $ & $ 0.0257 \pm 0.0006$ &F\\ 
Spectral type & M3 & M3 & F\\
Age (Gyr) & $> 0.8$ & $>0.8$ & F\\
\hline\hline
\end{tabular}
\tablebib{
\small
A) {\it Gaia} DR3 \citep{GaiaColl2023}. B) \citet{bailer_jones2021}.
C) {\it TESS} Input Catalogue Version 8 (TICv8, \citealt{Stassun2018}).
D) Two Micron All Sky Survey (2MASS, \citealt{Cutri2003}).
E) {\it Wide-field Infrared Survey Explorer} \citep[{\it AllWISE};][]{cutri_allwise}. F) This work. \\
\tablefoottext{a}{Kinematic velocity components adopting a right-handed, heliocentric Galactic system.}}
\end{threeparttable}
\end{table*}

\subsection{Stellar parameters}\label{sec:star_properties}

For consistency among the \texttt{THIRSTEE} sample, we derived the stellar parameters of TOI-521 and TOI-912 as in \cite{lacedelli2024}. We used the high-resolution ESPRESSO spectra to derive the spectroscopic parameters with {\sc SteParSyn}\footnote{\url{https://github.com/hmtabernero/SteParSyn/}} \citep{tab22}. We combined the BT-Settl stellar atmospheric models \citep{all12}, and the Vienna atomic line database \citep[VALD3, see][]{rya15} together with the Turbospectrum-generated \citep{ple12} grid of synthetic spectra. 
We followed \cite{mar21} to select the set of \ion{Fe}{i}, \ion{Ti}{i} and TiO lines and molecular bands optimal for M dwarfs.
For TOI-521, we obtained $T_{\rm eff}$~$=$~3544~$\pm$~24~K, $\log{g}$~$=$~4.97~$\pm$~0.10, and [Fe/H]~$=$~$-0.12$~$\pm$~0.08~dex, while for TOI-912 the obtained parameters are $T_{\rm eff}$~$=$~3572~$\pm$~25~K, $\log{g}$~$=$~4.98~$\pm$~0.09, and [Fe/H]~$=$~$-0.19$~$\pm$~0.07~dex.
As the reported error bars correspond to the internal errors only, to obtain a more realistic uncertainty on the stellar temperature we summed a systematic component following \cite{mar21}. Therefore, the final adopted values for the temperatures are 3544~$\pm$~100~K and 3572~$\pm$~100~K, respectively, for TOI-521 and TOI-912.

Additionally, for TOI-521 we used the infra-red coverage of the IRD spectra to derive the elemental abundances of the star.
We used the IRD template spectra after removal of the telluric features (tellurics are removed through the forward-modelling technique detailed in \citealt{Hirano20}) to obtain temperature, metallicity and the elemental abundances of Na, Mg, Ca, Ti, Cr, Mn, Fe, and Sr from the corresponding atomic lines, following \cite{Ishikawa20, Ishikawa22}. 
See \citealt{ikuta2025} for more details on the procedure. In the analysis, the microturbulent velocity was fixed at 0.5$\pm$0.5 km s$^{-1}$ for simplicity. 
We obtained $T_{\rm eff}$ = 3578 $\pm$ 101 K and $\lbrack{\rm Fe/H}\rbrack$ = 0.06 $\pm$ 0.13 dex, which is consistent within 1.2$\sigma$ with the values derived from the ESPRESSO dataset. The derived elemental abundances are reported in Table~\ref{table:star_IRD}.

\begin{table}[h]
\centering
\small
\caption{Stellar abundances of TOI-521 derived from IRD spectra.}
\begin{tabular}{lc}
\hline\hline
Parameter (dex) & Value\\
\hline
$\lbrack{\rm Na/H}\rbrack$  & 0.05 $\pm$ 0.10  \\
$\lbrack{\rm Mg/H}\rbrack$  & 0.11 $\pm$ 0.18  \\
$\lbrack{\rm Ca/H}\rbrack$ & 0.05 $\pm$ 0.11   \\
$\lbrack{\rm Ti/H}\rbrack$ & 0.35 $\pm$ 0.22   \\
$\lbrack{\rm Cr/H}\rbrack$ & 0.09 $\pm$ 0.13  \\
$\lbrack{\rm Mn/H}\rbrack$ & 0.14 $\pm$ 0.17  \\
$\lbrack{\rm Fe/H}\rbrack$ &0.06 $\pm$ 0.13   \\
$\lbrack{\rm Sr/H}\rbrack$ & 0.16 $\pm$ 0.19  \\
$\lbrack{\rm M/H}\rbrack$ & 0.09 $\pm$ 0.05  \\
\hline
\end{tabular}
\label{table:star_IRD}
\end{table}

Using the spectroscopic temperature obtained from the ESPRESSO dataset, we then derived the luminosity, mass, and radius of the two stars.  We initially computed the spectral energy distribution (SED; Fig~\ref{fig:sed}), with the same procedure detailed in \cite{lacedelli2025}.
The stellar bolometric luminosity is then obtained by integrating the SED, and used together with our derived \teff\ to infer the stellar radius through the Stefan-Boltzmann law. We obtained $R_{\star} = 0.421 \pm 0.028$~\rsun\ and $R_{\star} = 0.419 \pm 0.028$~\rsun\ for TOI-521 and TOI-912, respectively.
The stellar mass is then derived using the mass-radius relation of \cite{schweitzer2019}, leading to $M_{\star} = 0.420 \pm 0.030$~\msun\ and $M_{\star} = 0.418 \pm 0.030$~\msun\, respectively.
The surface gravity derived from our mass and radius values (\logg$= 4.81 \pm 0.09$ and \logg$= 4.82 \pm 0.09$, respectively) is consistent with the value derived from the spectroscopic analysis, probing the consistency of the results.
Interestingly, the two stars are very similar in terms of temperature, metallicity, mass and radius. 
They both lie on the main sequence of M dwarfs in the colour-magnitude diagram, that is, they have the same metallicity of the majority of M dwarfs in the solar neighbourhood (Fig.~\ref{fig:m_stars}).

Ultimately, we performed a kinematic analysis for both stars to obtain the Galactic space velocity components, from which we inferred that both stars likely belong to the thin disk 
\citep{bensby2014}. Both stars have space velocities that do not overlap with any of the known young star clusters of nearby stellar moving groups. Therefore, their kinematical age is $>$ 0.8 Gyr, which is consistent with the location of the star in the colour-magnitude diagrams (Fig.~\ref{fig:m_stars}). For TOI-912, this is in agreement with the gyrochronological age determined using our inferred rotational period (Sect.~\ref{sec:star_activity}), using various period-age relations ($t = 2.3 \pm  1.2$~Gyr from \citealt{barnes2007}; $t = 4.2 \pm  3.6$~Gyr from \citealt{mamajek2008}; $t = 5.4 \pm  4.2$~Gyr from \citealt{angus2015}). 
All the inferred stellar parameters are listed in Table~\ref{table:star_params}.

\subsection{Activity indices and rotational period}\label{sec:star_activity}
To investigate the stellar rotation period of both stars, we searched for periodic signals by computing the generalised Lomb-Scargle (GLS, \citealt{Zechmeister2009}) periodogram on the \textit{TESS} SAP\footnote{We analysed the SAP photometry to search for the rotational period of the star as the PDCSAP flux is automatically corrected for long-term trends and variability which could remove part of the rotational signal \citep{smith2012, Stumpe2012, Stumpe2014}.} \citep{twicken2010, morris2020} and long-term photometry, as well as various spectroscopic activity indicators. 
For the HARPS and ESPRESSO dataset, we used the activity indicators produced by the DRS, namely the bisector span (BIS), the full-width-half-maximum (FWHM), and the contrast of the cross-correlation function. 
For TOI-912, we additionally included for the ESPRESSO dataset the activity indexes provided by the \texttt{SERVAL} extraction, that is, the chromatic index (CRX), the differential line width (dLW), the Na \textsc{i} doublet, and the \halpha\ index, and we further computed the $S$-index using the \texttt{ACTIN2}\footnote{\url{https://github.com/gomesdasilva/ACTIN2}} code \citep{Gomes2018, Gomes2021}). 
For the IRD dataset, we computed the FWHM, the dLW, the CRX, and the line asymmetry (dV) following \cite{harakawa2022}.

\begin{figure}[h!]
\centering
  \includegraphics[width=\linewidth]{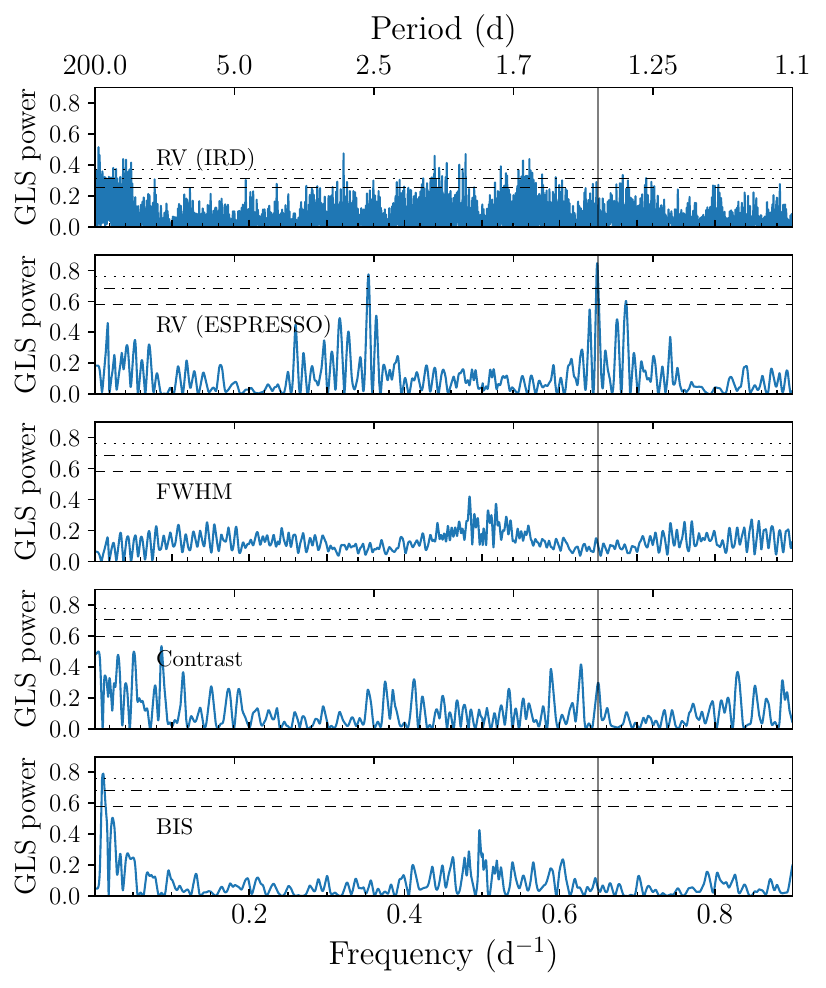}
  \caption{GLS periodogram of IRD and ESPRESSO RVs, and ESPRESSO activity indicators of TOI-521. The vertical gray shows the planetary period of TOI-521.01 at $\sim 1.54$. In the ESPRESSO RVs periodogram, the significant peak at $f = 0.351$~d$^{-1}$ ($\sim 2.85$~d) corresponds to the daily alias of the planetary period. The horizontal dashed, dash-dotted, and dotted horizontal lines show the $10\%$, $1$\%, and 0.1\% FAP levels, respectively. 
  }
    \label{fig:TOI521_periodogram}
\end{figure}

For TOI-521, no clear and unique periodicity emerges from the ASAS-SN, ZTF or \textit{TESS} SAP photometry (see Fig.~\ref{fig:asas_sn_toi521}). Similarly, the spectral activity indicators do not show evidence of a common clear, significant peak which could be related to stellar activity (Fig.~\ref{fig:TOI521_periodogram}), a part from the bisector span, which shows a long-period peak at $\sim 86$~d. On the other side, the signal corresponding to the planetary candidate TOI-521.01 is clearly identified at $\sim 1.54$ days in the ESPRESSO RVs, even though not present in the IRD RVs due to the higher internal errors and scatter of the dataset (Fig.~\ref{fig:TOI521_IRD_HARPS_periodogram}).
The absence of a clear rotational period for TOI-521 is not surprising considering the low activity level of the star (\logRHK~$= -5.500 \pm  0.036$). Considering the derived \logRHK, the star is probably a slow rotator, with a predicted rotational period of $P_{\rm rot} = 83 \pm 17$~d, as derived from the activity-period relations of \cite{suarez_mascareno2016} for M dwarfs. Such a prediction is compatible with the peak identified in the BIS periodogram.

\begin{figure}[h!]
\centering
  \includegraphics[width=\linewidth]{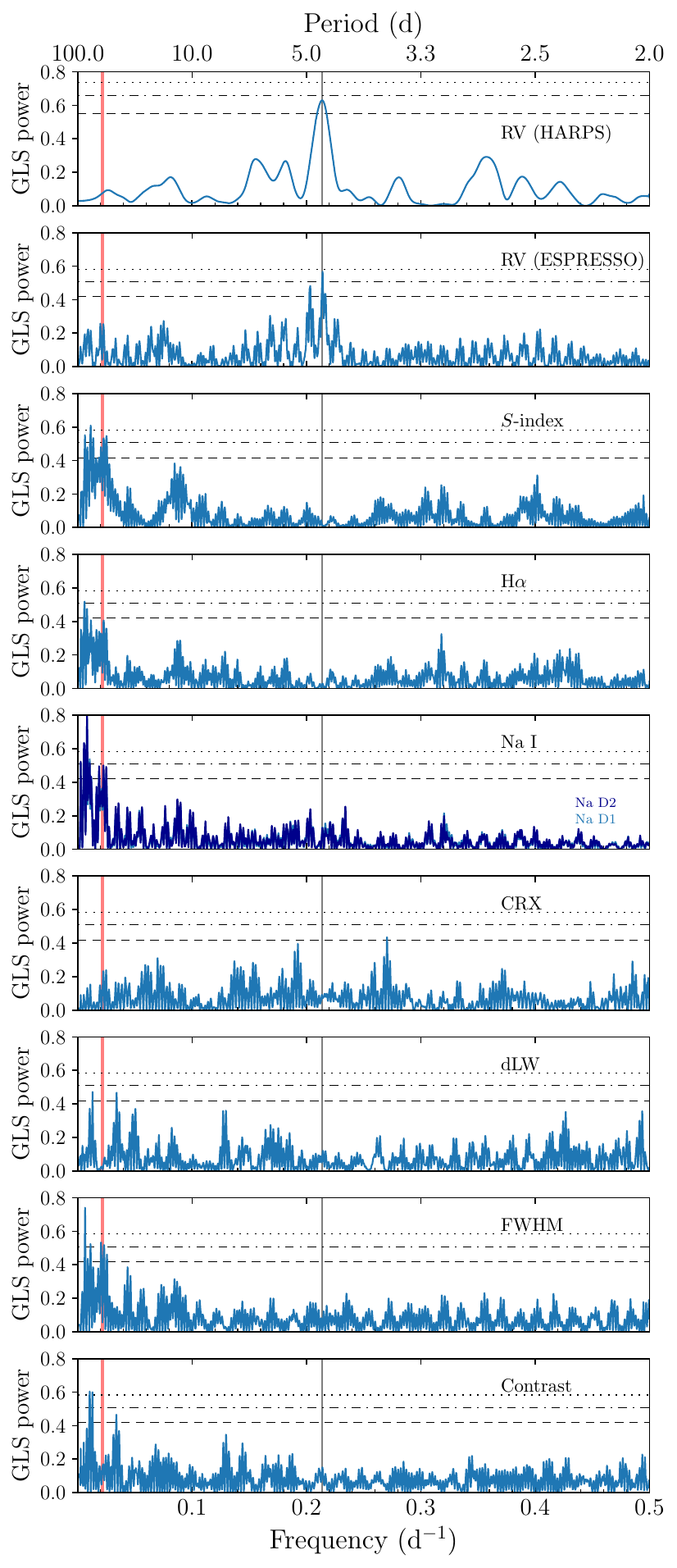}
  \caption{Same as Fig.~\ref{fig:TOI521_periodogram}, but for the HARPS and ESPRESSO RVs, and the ESPRESSO activity indicators of TOI-912. The vertical gray and red lines show the planetary signal of TOI-912.01 ($\sim 4.7$ d) and the stellar rotational period ($\sim 46$ d).
  }
    \label{fig:TOI912_periodogram}
\end{figure}

For TOI-912, the ASAS-SN photometry shows a significant peak around $\sim 48$~d, both in the $V$ and $g$ band. Similarly, the \textit{TESS} SAP photometry shows a broad peak around $40-50$~d (Figure~\ref{fig:asas_sn_TOI912}). 
This is also in line with the rotational period computed from the activity-period relations for a  \logRHK~$= -5.176 \pm  0.011$, resulting in $P_{\rm rot} = 47 \pm 5$~d. 
Stellar rotation with such periodicity is further supported by the spectroscopic activity indicators analysis. In fact, as Figure~\ref{fig:TOI912_periodogram} shows, many of the ESPRESSO activity indicators manifest the presence of a peak around $\sim 46$~d, which correspond to the rotational period of the star as identified in the photometry. This peak is particularly evident in the $S$-index time series, which we therefore used as a proxy of the stellar activity, and model it together with the RVs to better inform the activity model (see Sect.~\ref{sec:rv_fit_only}). Moreover, the signal at $\sim 45$~d is also identified in the ESPRESSO RVs when performing an iterative periodogram search removing a sinusoidal model with the period of the most significant peak (see Fig.~\ref{fig:periodogram_iterative}). Due to the lower number of points, the higher scatter, and the lower precision of the HARPS dataset, the stellar rotational period is not identified in the HARPS RVs, nor in the corresponding activity indicators (see Figure~\ref{fig:TOI521_IRD_HARPS_periodogram}). However, as Fig.~\ref{fig:TOI912_periodogram} shows, even if no stellar rotational period is identified, the most significant peak in the HARPS RVs periodogram (as well as in the ESPRESSO one) is identified at $\sim 4.7$ days, corresponding to the planetary candidate TOI-912.01.

\section{Data analysis and results}\label{sec:global_analysis}
  
\subsection{Photometric fit}\label{sec:photometric_fit_only}
We performed a photometric fit including all the available data with the \pyorbit\footnote{\url{https://github.com/LucaMalavolta/PyORBIT}.} code \citep{Malavolta2016, Malavolta2018}. 
We detrended the \textit{TESS} light curves using a biweight time-windowed slider ($1$-d window) as implemented in \texttt{WOTAN}\footnote{\url{https://github.com/hippke/wotan}.} \citep{hippke2019}, masking the in-transit points to avoid affecting the transit shape\footnote{We compared the results of the detrended data with a fit including a Gaussian process (GP) regression within the model, using a Matérn-3/2 kernel. Given the consistency of the obtained planetary parameters and uncertainties, we decided to adopt the pre-detrended light curves in the global fit to reduce the computational time.}. 
For each ground-based transit, we included a 2$^{\rm nd}$-order polynomial to model the out-of-transit variability. 
We fitted the central time of transit ($T_0$), period ($P$), stellar radius ratio ($R_\mathrm{p}/$\rstar), impact parameter ($b$), stellar density (\rhostar), adopting a Gaussian prior as derived from the stellar analysis, and limb-darkening (LD) coefficients ($q_1$, $q_2$) as parametrised in \cite{Kipping2013}. For each bandpass, we computed the LD coefficients with the \texttt{PyLDTk}\footnote{\url{https://github.com/hpparvi/ldtk}.} code \citep{Husser2013, Parviainen2015},  and we adopted them as Gaussian priors with a custom $0.1$ uncertainty. 
For each light curve, we also included a jitter term to account for uncorrelated noise. 
For all fitted parameters we adopted uniform priors, if not specified otherwise. 
We explored the parameter space with the \pyorbit\ \texttt{PyDE}\footnote{\url{https://github.com/hpparvi/PyDE}} + {\tt emcee} \citep{ForemanMackey2013} configuration, using the same model specifications and convergence criteria described in \cite{lacedelli2024}.

\subsection{RV fit}\label{sec:rv_fit_only}

We fitted all the available RVs, testing for each planet with both a circular and an eccentric model with free eccentricity ($e$) \footnote{We adopted \cite{Eastman2013} parametrisation [$\sqrt{e} \sin{\omega}$, $\sqrt{e} \cos{\omega}$], with $\omega =$ argument of periastron.}, comparing the Bayesian evidence ($\mathcal{Z}$) through the \texttt{dynesty} nested sampling algorithm \citep{skilling2004, skilling2006, speagle2020} as implemented in \pyorbit, adopting a sampling efficiency of 0.3 and 1000 live points. We fitted the RV semi-amplitude ($K \in [0.01$-$100$]~\ms), the $T_0$ and $P$ (with narrow, uniform priors around the values obtained from the photometric fit), a systemic RV offset ($\gamma$), and a jitter term ($\sigma_{\rm j}$).

\begin{figure}[h!]
\centering
  \includegraphics[width=\linewidth]{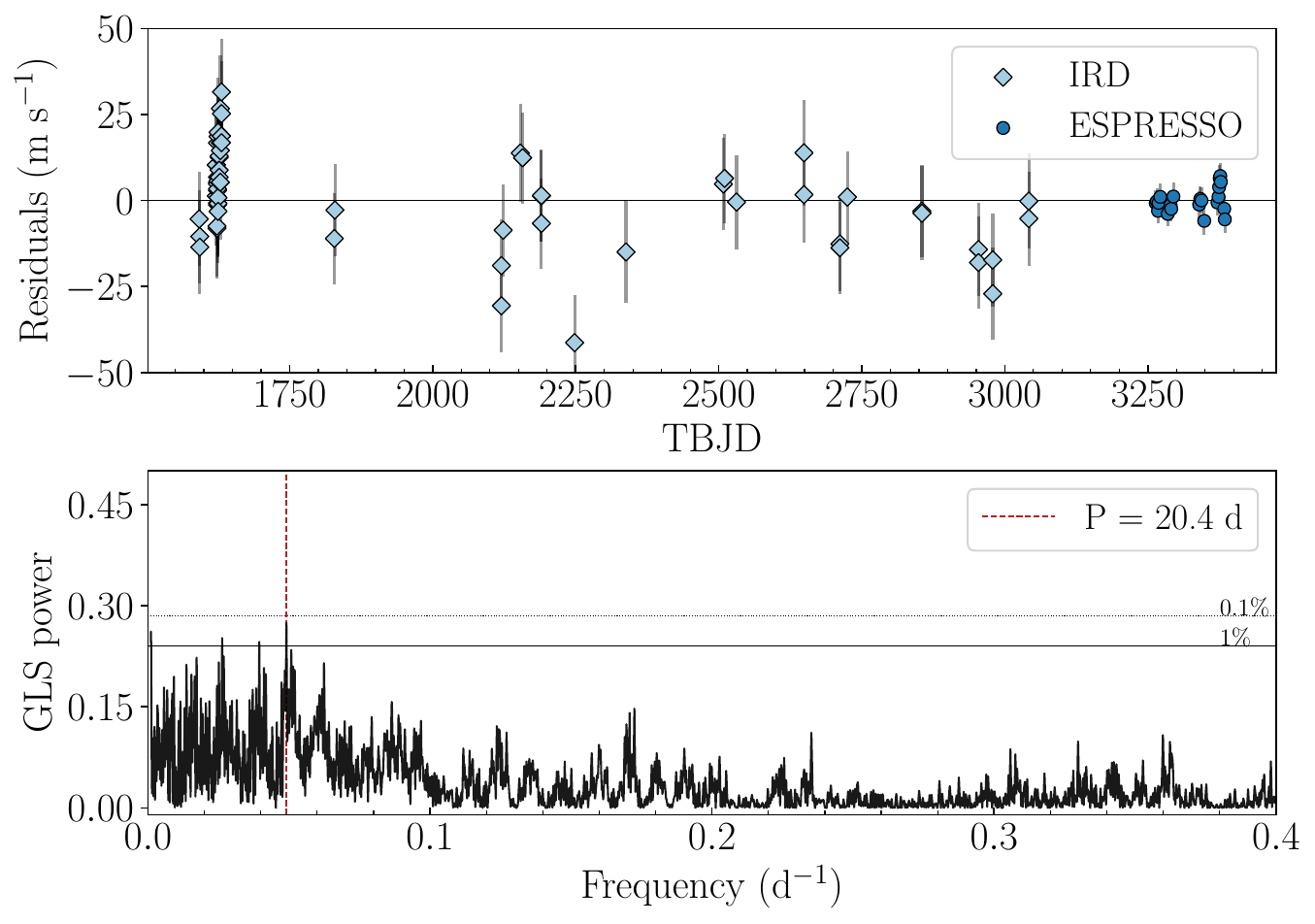}
  \caption{RV residuals and GLS periodogram after the 1-planet (circular) model fit for TOI-521. The vertical dashed line in the periodogram highlights the most significant peak.
  }
    \label{fig:TOI521_residuals}
\end{figure}

For TOI-512 b, the eccentric model was not favoured by the Bayesian evidence, with $\Delta \ln\mathcal{Z} = 0.9$ \citep{kass&raftey1995}, and the derived eccentricity was unconstrained, and in any case consistent with $0$ within $1\sigma$. We therefore assumed the simplest model with a circular orbit. However, the periodogram of the residuals after the 1-planet fit showed a significant peak at $\sim 20$ d (Fig.~\ref{fig:TOI521_residuals}). Considering also the high jitter of ESPRESSO ($\sigma_{\rm j} \sim 3.4$~\ms), we therefore tested a 2-planet model, leaving the period of the second candidate free to span between 4 and 200 days. The 2-planet model identifies a second signal at $P \simeq 20.3$ days, with a semi-amplitude of $3.9 \pm 0.9$~\ms. This model is favoured by the Bayesian evidence 
with respect to the 1-planet model,
with a difference of $\Delta \ln\mathcal{Z} = 2.4$. The signal at $\sim 20$ days is not identified in any of the activity indicators, and it is inconsistent with the long rotational period suggested from the \logRHK\ (Sect.~\ref{sec:star_activity}). We therefore associate it with a second candidate in the system, even though further observations will be needed to finally confirm it as a planet, given that the model is statistically favoured, but not strongly. 
We tested both a circular and an eccentric model for the additional candidate, and we found that the eccentric model was not favoured by the Bayesian evidence, and returned an unconstrained eccentricity, consistent with $0$. We therefore assumed as a final model a 2-planet system with both planets in circular orbits. We note that adding a second planet in the model did not influence the recovered semi-amplitude of TOI-521 b, which was consistent among the two fits ($K = 5.9_{-1.3}^{+1.2}$ and $ 5.7_{-1.0}^{+0.9} $~\ms, respectively).

For TOI-912, given the results of our frequency analysis (Sect.~\ref{sec:star_activity}), we also included in each fit a GP regression model with a quasi-periodic kernel \citep{Grunblatt2015} to model the stellar activity, fitting the stellar rotation period (\prot), the characteristics decay timescale ($P_\mathrm{dec}$), the coherence scale ({\it w}), and the GP amplitude ($A$) as hyper-parameters. We modelled together the RVs and the $S$-index time series, in order to better inform the GP \citep{Rajpaul2015, barragan2023}. 
For this planet, the eccentric model ($\ln\mathcal{Z} = -165.7 \pm 0.2$) is strongly favoured by the Bayesian evidence with respect to the circular one ($\ln\mathcal{Z} = -170.6 \pm  0.2$), with $\Delta\ln\mathcal{Z} = 4.9 $ \citep{kass&raftey1995}.
However, given the low number of RV points and the spread of the data, which are not optimally sampling the rotation period of the star, we also tested a simpler model fitting the activity with a sinusoidal, to check if the GP model is biasing the eccentricity. 
In the 2-Keplerian model, the eccentricity is consistent with the value obtained from the GP fit ($e \sim 0.50 \pm 0.05$). 
Such high eccentricity is also obtained from a simpler 1-planet eccentric model without GP that we tested, even though with less significance ($e = 0.34_{-0.16}^{+0.14}$). However, in this 1-planet model the RV jitter is unusually high, especially for the ESPRESSO dataset ($\sigma_{\rm j, ESPRESSO} = 2.1$~\ms, $\sigma_{\rm j, HARPS} = 1.1$~\ms), hinting at the presence of an additional signal, as expected. This is also reflected in the Bayesian evidence comparison. In fact, the models including a second signal to account for the activity are both favoured, even though only slightly when using a Keplerian ($\Delta \ln\mathcal{Z} = 1.6$), while having stronger evidence ($\Delta\ln\mathcal{Z} = 4.9 $) when using a GP regression. This could be related to the fact that in the 2-Keplerian model the period of the second Keplerian is not well defined, showing various peaks in the $43-50$~d range. The same happens when applying an uninformed GP only on the RV time series, while such a degeneracy is not present when using a GP model informed with the $S$-index, which results in a very defined peak at $\sim 46$~d.
Finally, apart from the uniform prior used in all the previous fits, we also tested different priors for the eccentricity, namely a half-Gaussian zero-mean prior \citep{vanEylen2019}, a $\beta$ function \citep{Kipping2013}, and a truncated Rayleigh function (with $\sigma = 0.2$). While all these priors favour low eccentricity, in the GP model the results do not change with respect to the uniform prior when using the $\beta$ and the Rayleigh function, while the half-Gaussian prior leads to a less significant eccentricity ($e = 0.21 \pm 0.14$), still higher than the initial prior. The same happens with the 2-Keplerian model, even though in this case the eccentricity is less significant also for the other two functions, still pointing towards higher values with respect to the initial prior ($e = 0.37_{-0.21}^{+0.14}$ and $0.15_{-0.08}^{+0.09}$, respectively).

Considering this, and based on the Bayesian evidence, we decided to adopt the 1-planet model with free eccentricity and GP modelling informed with the $S$-index in the following global fit. However, we notice that the results on the eccentricity are somehow dependent on the choice of priors, and that the eccentric model is supported, but not overwhelmingly confirmed, by the Bayesian evidence. More RV observations, possibly optimised to sample the rotational period of the star, will be needed to confirm it. Nonetheless, the recovered semi-amplitude of TOI-912 b was consistent among all models, probing the robustness of the semi-amplitude detection. No additional significant peaks are identified in the periodogram of the residuals after our global fit (see Figure~\ref{fig:global_rv}, bottom panel). 

\subsection{Joint photometric and RV modelling}\label{sec:joint_fit}

We performed a joint modelling with \pyorbit\ of all photometric and RV data, fitting the parameters as described in Sect.~\ref{sec:photometric_fit_only} and \ref{sec:rv_fit_only}. We assumed for each planet as final model the one derived in Sect.~\ref{sec:rv_fit_only}.

\begin{figure*}[h!]
\centering
  \includegraphics[width=0.485\linewidth]{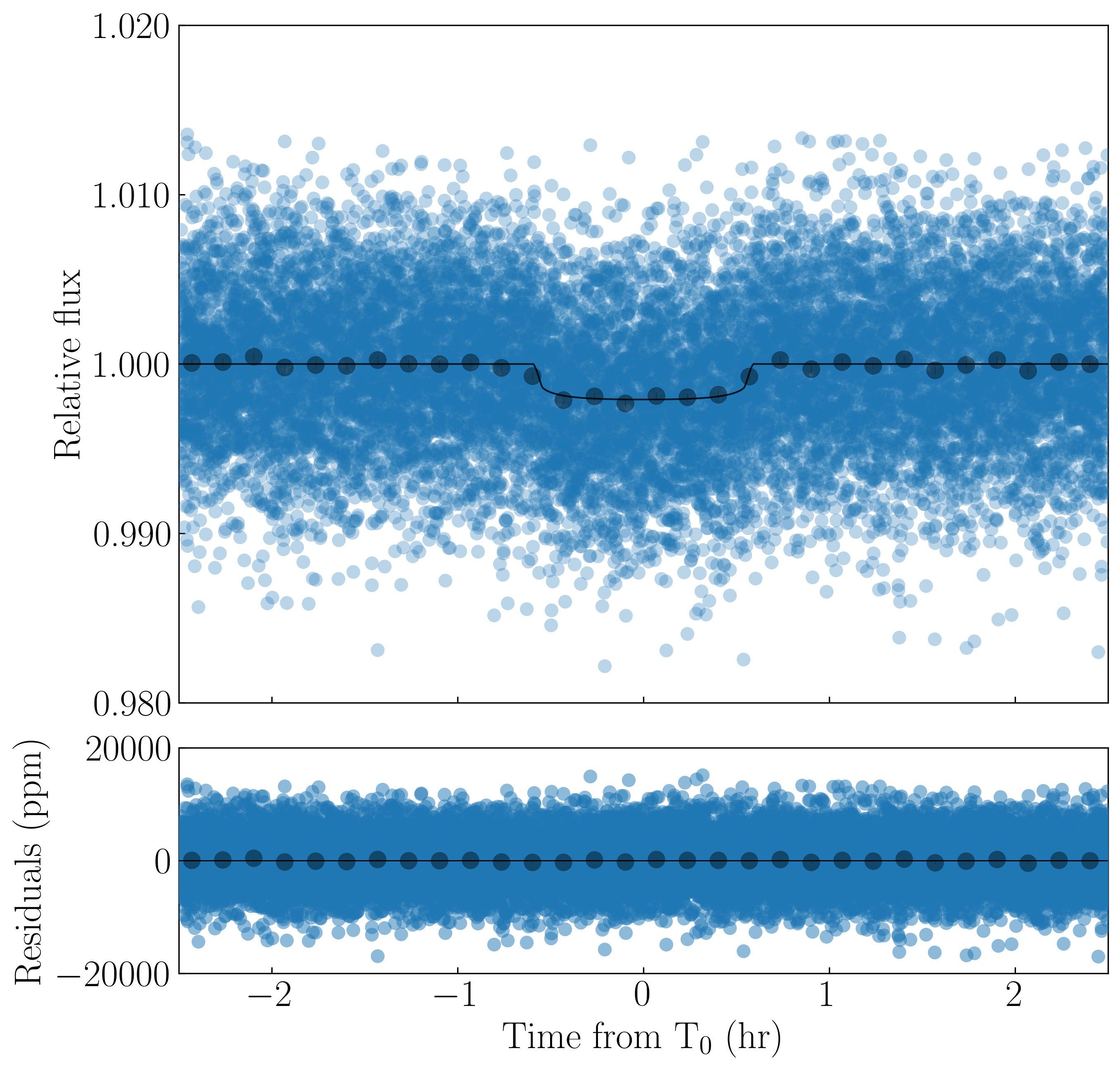}
  \includegraphics[width=0.465\linewidth]{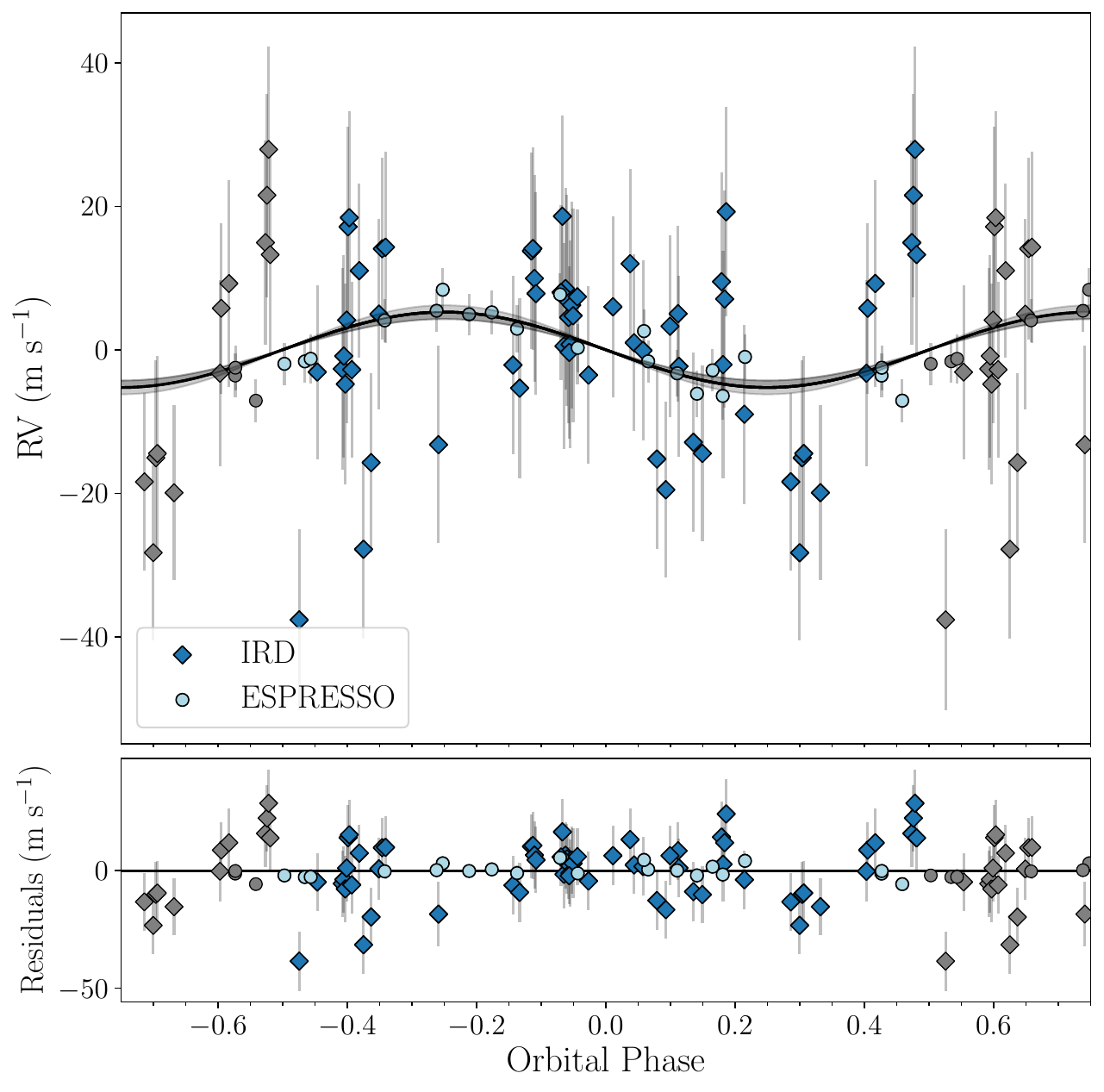}
  \caption{Phase-folded \tess\ light curve (left) and RVs (right) of TOI-521 b from the joint photometric and RV analysis.The bottom panels show the model residuals. The best-fitting model is plotted as a solid black line. In the transit plot, the black dots show the data binned over $15$~min. In the RV plot, the error bars include the jitter term, added in quadrature to the nominal error, and the grey shaded area highlights the $\pm 1 \sigma$ region. 
  }
    \label{fig:phase_plot_TOI521}
\end{figure*}

\begin{figure}[h!]
\centering
  \includegraphics[width=\linewidth]{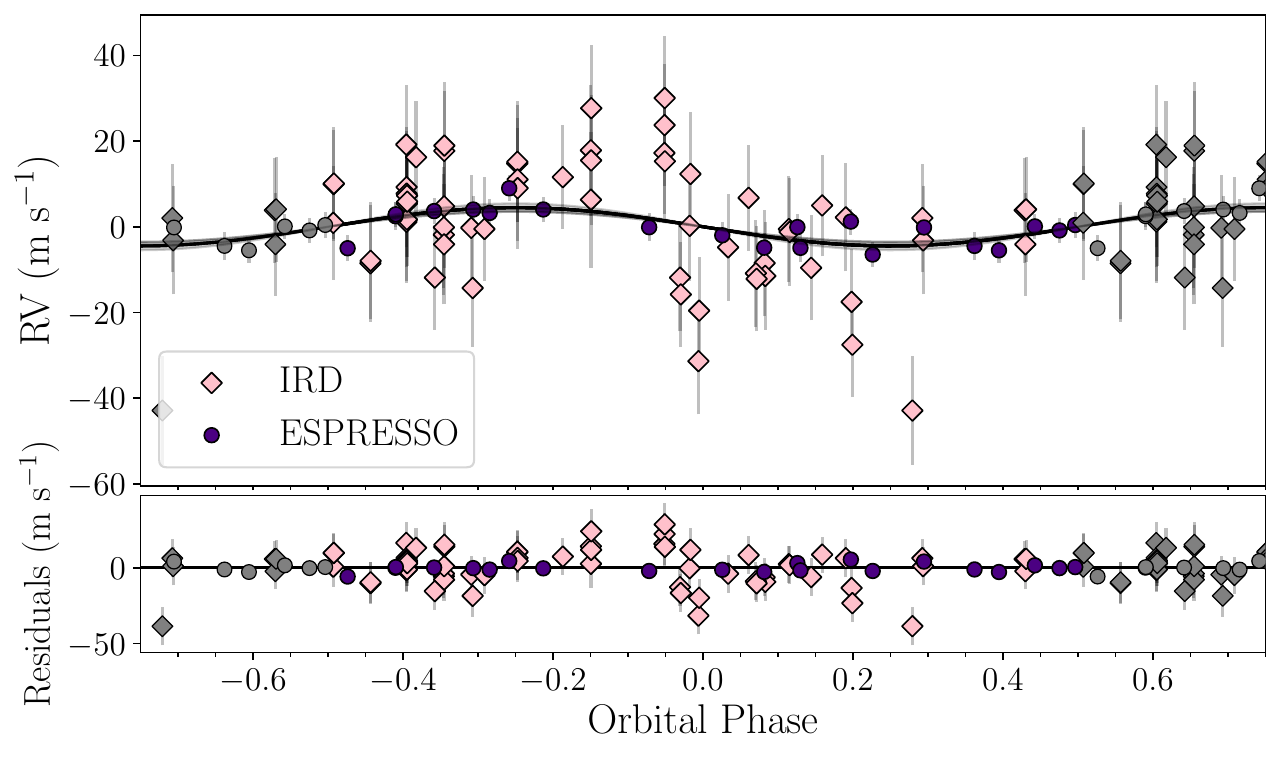}
  \caption{Same as Fig.~\ref{fig:phase_plot_TOI521}, but showing the phase-folded RVs of the candidate TOI-521 c.
  }
    \label{fig:phase_plot_TOI521c}
\end{figure}

\begin{figure*}[h!]
\centering
  \includegraphics[width=0.43\linewidth]{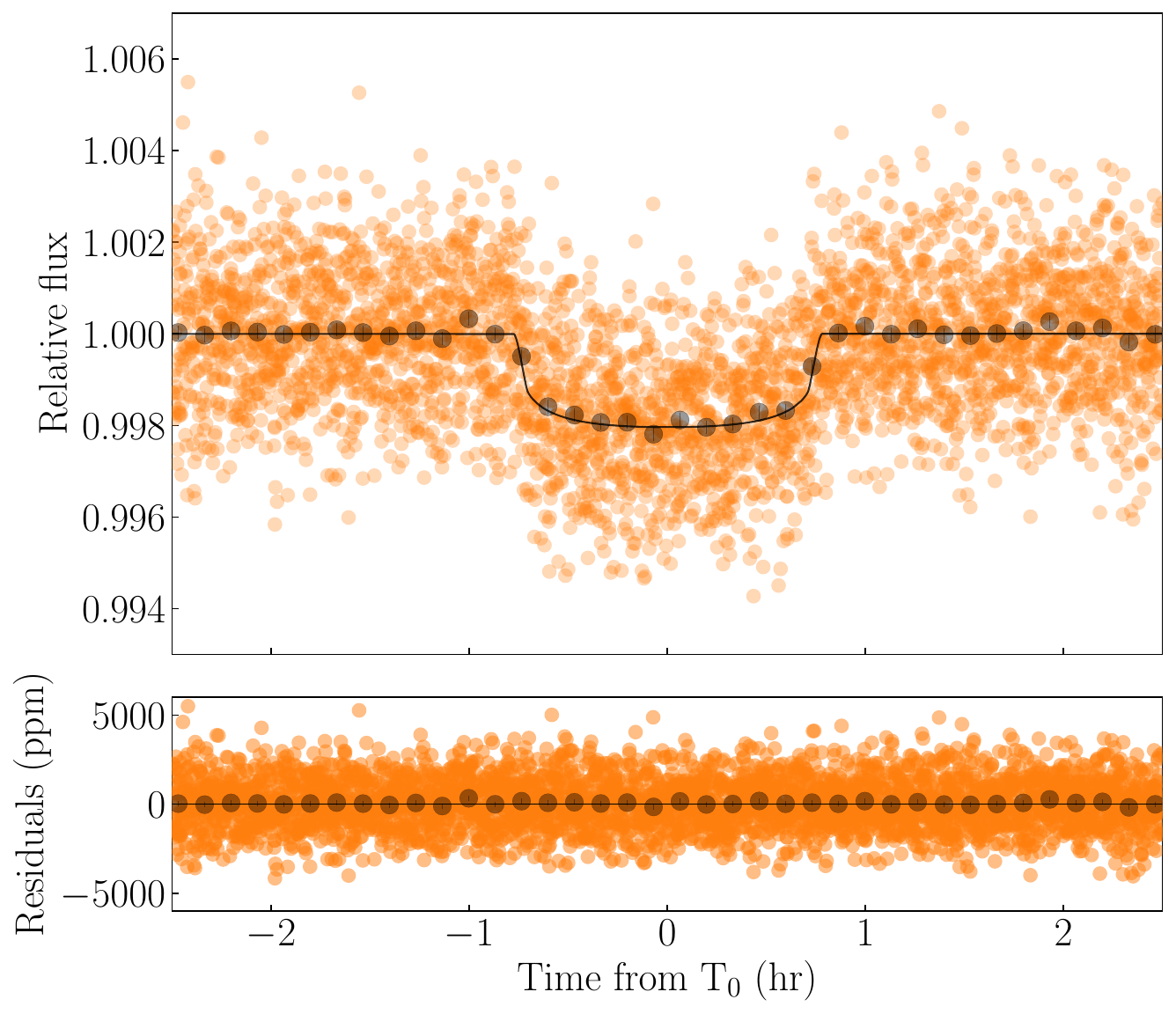}
  \includegraphics[width=0.46\linewidth]  {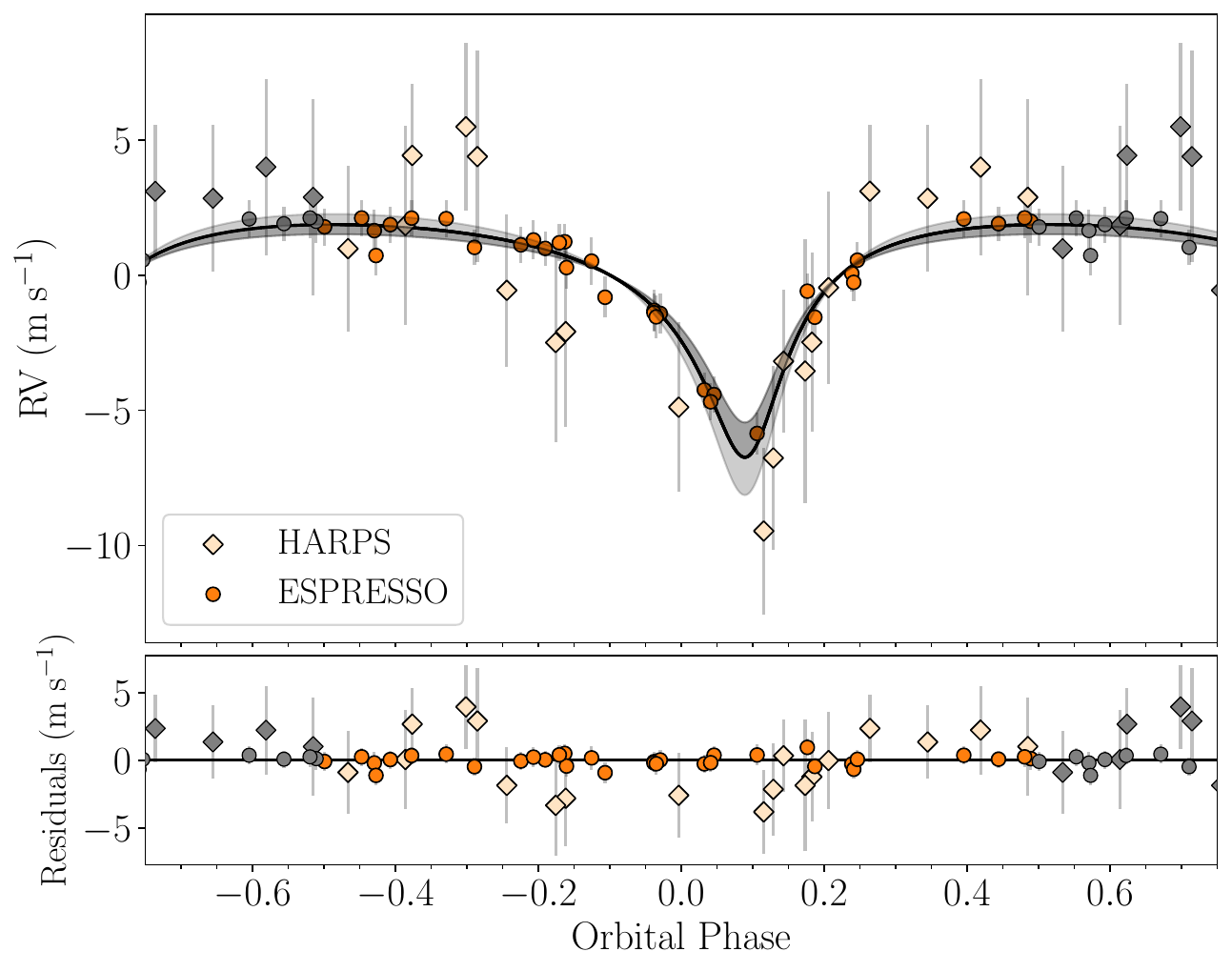}
  \caption{Same as Fig.~\ref{fig:phase_plot_TOI521}, but for TOI-912 b. 
  }
    \label{fig:phase_plot_TOI912}
\end{figure*}

We report in Table~\ref{table:joint_parameters} the final parameters derived for the two systems under analysis. Figures \ref{fig:phase_plot_TOI521} and \ref{fig:phase_plot_TOI521c} report the phase-folded \textit{TESS} light curve and RVs for the two planets in the TOI-521 system, and Fig. \ref{fig:phase_plot_TOI912} shows the results for TOI-912 b. The ground-based transit models and the global RV models are shown in Figs.~\ref{fig:ground_based} and \ref{fig:global_rv}, respectively. 

For TOI-521 b, we inferred a radius of $1.98 \pm 0.14$~\rearth\ and a mass of $5.3 \pm 1.0$~\mearth\ (from $K = 5.26_{-0.98}^{+0.95}$~\ms), implying a bulk density of $3.8 \pm 1.1$~\gcm. A very similar density ($4.0 \pm 0.9$~\gcm) is recovered for TOI-912 b, which has an almost identical radius \rplanet\ = $1.93 \pm 0.13$~\rearth\ and \mplanet\ = $5.1 \pm 0.5$~\mearth\ (from $K = 4.29_{-0.37}^{+0.39}$~\ms) as TOI-521 b. 
For this planet, our model shows a significant eccentricity of $e = 0.58 \pm 0.02$. From our fit, we recovered a rotational period for TOI-912 of $45.9_{-0.2}^{+0.6}$ d, consistent with our stellar analysis (Sect.~\ref{sec:star_activity}).
Both planets are orbiting at short distance from the host star, with $P \sim 1.54 $~d (TOI-521 b) and $\sim 4.67$ d (TOI-912 b), implying equilibrium temperatures of \teq $= 794 \pm 35$~K and $551 \pm 27$~K, respectively. 
Moreover, we identify an additional signal in the TOI-521 system at $P = 20.3$~d, which we associate with a second, possibly non-transiting candidate (see Sect.~\ref{sec:matrix}) with a minimum mass of $M_{\rm p} \sin{i} = 10.7_{-2.4}^{+2.5}$~\mearth\ (from $K = 4.5 \pm 1.0$~\ms). 
Due to the presence of this second candidate, we performed a transit time variation (TTV) analysis, to check for possible mutual interactions between the two. Due to its unusually high eccentricity, we performed the same analysis for TOI-912 b, to investigate the presence of additional, non-detected interacting planets in the system. 
We performed the fit as described in Sect.~\ref{sec:photometric_fit_only}, but leaving each $T_0$ free to vary\footnote{For the epochs having multiple transits, i.e. due to multi-colour simultaneous photometry with MuSCAT2, we included in the fit a unique transit, selecting the one showing the lowest RMS.}, and using the derived eccentricity value for TOI-912 b as a Gaussian prior. As shown in Fig.~\ref{fig:ttv}, when comparing the observed $T_0$ with the ones calculated from the linear ephemeris (Table~\ref{table:joint_parameters}), we see no significant TTV patterns, with all the computed transit times consistent with the linear ephemeris at 2$\sigma$.

\begin{table*}
\renewcommand*{\arraystretch}{1.2}
\centering
\caption{Best-fitting parameters of two systems under analysis.}
\label{table:joint_parameters} 
\small
\begin{threeparttable}[t]
\centering
\begin{tabular}{l c c c} 
  \hline\hline       
 \multicolumn{4}{c}{Planetary parameters}\\
 \hline 
   & TOI-521 b & TOI-521 c$^*$ & TOI-912 b \\
  \hline
  $P$ (d) & $ 1.54285047_{-0.0000008}^{+0.0000010} $ & $20.35 \pm 0.02$ & $4.6780691 \pm 0.0000017$ \\
  $T_0$ (TBJD)$^a$  &  $1492.7153 \pm 0.0005$ & $1489.3_{-1.7}^{+1.5}$ & $1628.4061 \pm 0.0004$\\
  $a/$\rstar & $10.1_{-0.3}^{+0.2}$  & $56.3_{-1.7}^{+1.3}$ &  $20.7_{-1.2}^{+1.1}$  \\
  $a$ (AU)   & $0.0195 \pm 0.0003$ & $0.109 \pm 0.002$ &$0.041\pm 0.001$ \\
  $R_\mathrm{p}/$\rstar & $0.0431 \pm 0.0009$ & -&$0.0421_{-0.0007}^{+0.0008}$\\
  \rplanet\ (\rearth) & $1.98 \pm 0.14$  & -& $1.93 \pm 0.13$ \\
  $b$ & $0.24_{-0.14}^{+0.11}$ & - &$0.34_{-0.22}^{+0.18}$ \\
  $i$ (deg) & $88.6_{-0.7}^{+0.8}$ & -&$88.7_{-0.6}^{+0.8}$ \\
  $T_{14}$ (h) & $1.18 \pm 0.01$ & - &$1.7 \pm 0.1$ \\
  $e$ & 0 (fixed) & 0 (fixed) &  $0.58 \pm 0.02$ \\
  $\omega$ (deg)   & $90$ (fixed)  & $90$ (fixed) & $-169_{-6}^{+9}$ \\
  $K$ (\ms)  & $ 5.26_{-0.98}^{+0.95}$ & $4.5 \pm 1.0$& $4.29_{-0.37}^{+0.39}$\\
  \mplanet\ (\mearth) & $5.3 \pm 1.0$  & - & $5.1 \pm 0.5$\\
  \mplanet\ $\sin{i}$\ (\mearth) & -  & $10.7_{-2.4}^{+2.5}$& -\\
  \rhoplanet\ (\gcm)  & $3.8 \pm 1.1$ & - & $4.0 \pm 0.9$\\
  $\rho / \rho_{\oplus}$  & $ 0.68 \pm 0.19$ & - &$0.72 \pm 0.16$\\
  $S_{\rm p}$ ($S_{\oplus}$)  & $66 \pm 12$ & $2.1 \pm 0.4$ & $15 \pm 3$\\
  $T_{\rm eq}^b$ (K)  & $794 \pm 35$ & $336 \pm 15$ &  $551 \pm 27$\\
  $g_{\rm p}^c$ (m s$^{-2}$)  &  $13 \pm 3$ & - & $13 \pm 2$ \\[1ex] 
 \hline 
 \multicolumn{4}{c}{Stellar parameters}\\
 \hline
  \rhostar\ (\rhosun) & \multicolumn{2}{c}{$5.8_{-0.5}^{+0.4}$ } & $5.5 \pm 0.5$ \\
  $u_{1, \mathrm{TESS}}$ & \multicolumn{2}{c}{$0.15_{-0.08}^{+0.09}$}& $0.21 \pm 0.08$\\
  $u_{2, \mathrm{TESS}}$ &  \multicolumn{2}{c}{$0.42 \pm 0.10$} & $0.45 \pm 0.09$ \\
  $u_{1, \mathrm{RG715}}$ & \multicolumn{2}{c}{-} & $0.19_{-0.09}^{+0.10}$ \\
  $u_{2, \mathrm{RG715}}$ & \multicolumn{2}{c}{-} & $0.26 \pm 0.10$ \\
  $u_{1, \mathrm{g'}}$ & \multicolumn{2}{c}{$0.25_{-0.10}^{+0.09}$} & $0.24 \pm 0.09$ \\
  $u_{2, \mathrm{g'}}$ & \multicolumn{2}{c}{$0.51_{-0.10}^{+0.09}$} & $0.47 \pm 0.10$ \\
  $u_{1, \mathrm{r'}}$ & \multicolumn{2}{c}{$0.23_{-0.14}^{+0.16}$} & $0.26 \pm 0.10$ \\
  $u_{2, \mathrm{r'}}$ & \multicolumn{2}{c}{$0.43_{-0.18}^{+0.17}$} & $0.34 \pm 0.10$ \\
  $u_{1, \mathrm{i'}}$ & \multicolumn{2}{c}{$0.20_{-0.12}^{+0.14}$} & $0.26 \pm 0.10$ \\
  $u_{2, \mathrm{i'}}$ & \multicolumn{2}{c}{$0.27_{-0.17}^{+0.18}$} & $0.34 \pm 0.10$ \\
  $u_{1, z'}$ & \multicolumn{2}{c}{$0.17 \pm 0.09$} & $0.18_{-0.09}^{+0.10}$ \\
  $u_{2, z'}$ & \multicolumn{2}{c}{$0.24 \pm 0.10$} & $0.23 \pm 0.10$ \\
  $u_{1, \mathrm{ExTrA}}$ & \multicolumn{2}{c}{-} & $0.11_{-0.07}^{+0.08}$\\
  $u_{2, \mathrm{ExTrA}}$ & \multicolumn{2}{c}{-} & $0.18 \pm 0.09$\\
  $\sigma_{\rm j, ESPRESSO}$ (\ms) &\multicolumn{2}{c}{$2.6_{-0.6}^{+0.7}$} & $0.5 \pm 0.3$ \\
  $\gamma_{\rm ESPRESSO}$ (\ms) &   \multicolumn{2}{c}{$ -9552.6\pm 0.7$} & $ 2.7 \pm 1.0$\\[1ex]
  $\sigma_{\rm j, HARPS}$ (\ms) &   \multicolumn{2}{c}{-} & $0.9_{-0.6}^{+0.9}$ \\
  $\gamma_{\rm HARPS}$ (\ms) &      \multicolumn{2}{c}{-} & $11032.4_{-0.9}^{+1.0} $  \\[1ex]
  $\sigma_{\rm j, IRD}$ (\ms) &     \multicolumn{2}{c}{$11_{-1}^{+2}$} & - \\
  $\gamma_{\rm IRD}$ (\ms) &        \multicolumn{2}{c}{$-2.8_{-1.7}^{+1.8}$ }& -  \\[1ex]
   \hline 
 \multicolumn{4}{c}{GP hyper-parameters}\\
 \hline
  $P_{\rm rot}$ (d) & \multicolumn{2}{c}{-} & $ 45.9_{-0.2}^{+0.6}$ \\
  $P_{\rm dec}$ (\ms) & \multicolumn{2}{c}{-} & $ 496_{-146}^{+160}$\\
  $w$ (\ms) & \multicolumn{2}{c}{-} & $0.16 \pm 0.01$ \\
  $A_{\rm ESPRESSO}$ (\ms) & \multicolumn{2}{c}{-} & $2.6_{-0.6}^{+0.7}$ \\
  $A_{\rm HARPS}$ (\ms) & \multicolumn{2}{c}{-} & $1.0_{-0.7}^{+1.2}$ \\
  $A_{\rm S-index}$ (\ms) &\multicolumn{2}{c}{-}  &  $0.8_{-0.1}^{+0.2}$\\[1ex]
 \hline
\end{tabular}
\tablefoot{\tablefoottext{$*$}{Planet candidate (see Sect.~\ref{sec:rv_fit_only}.} 
    \tablefoottext{a}{BJD$-2457000$.} 
\tablefoottext{b}{$T_{\rm eq} = T_\star \, \biggl(\dfrac{R_\star}{2a}\biggr)^{1/2} \, [f(1-A_{\rm B})]^{1/4}$, with $f=1$ and $A_{\rm B} = 0$.}  \tablefoottext{c}{Surface gravity.} 
}
\end{threeparttable}
\end{table*}

\subsection{Photometric search and sensitivity limits for TOI-521}\label{sec:matrix}

\begin{figure}[h!]
\centering
  \includegraphics[width=\columnwidth]{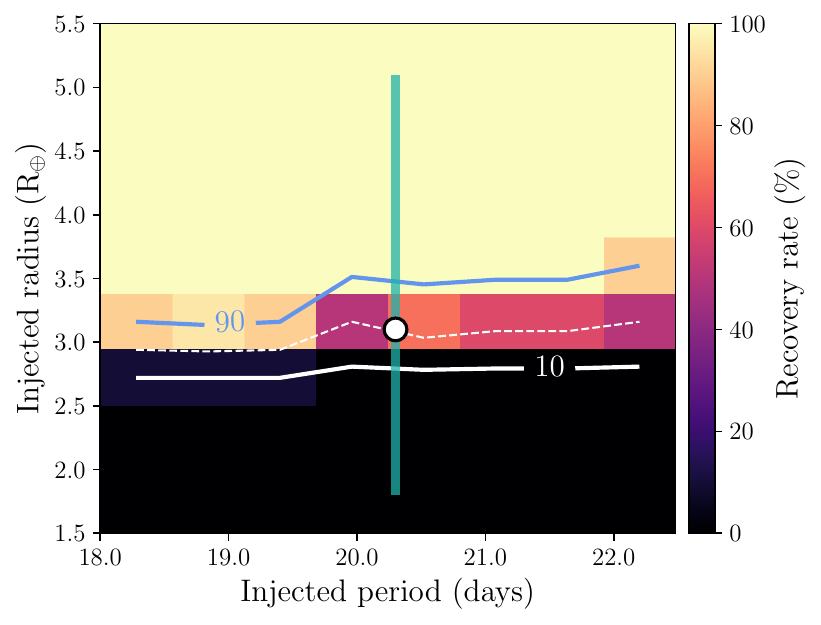}
  \caption{Results of the injection-recovery experiment performed with MATRIX to assess the detectability of the candidate TOI-521\,c. The colour scale represents recovery rates, where bright yellow indicates high recovery and dark purple/black indicates low recovery. The solid blue line marks the 90\% recovery contour, the dashed white line indicates the 50\%, and the solid white line shows the 10\%. The white dot marks the nominal value for TOI-521\,c as derived using \texttt{spright}, and the green solid line denotes the 95\% credible interval of the most probable radius distribution.}
  \label{fig:TOI521_IR}
\end{figure}

Our analysis shows the minimum mass of the additional candidate TOI-521 c is $10.7_{-2.4}^{+2.5}$~\mearth. With this mass, we used \texttt{spright} \citep{parviainen2024} to compute its most probable radius distribution, finding it likely has a radius of $3.1$~\rearth\ and a 95\% probability of falling between 1.8 and 5.1~\rearth. We searched for this planet in \textit{TESS} data using the SHERLOCK package \citep{pozuelos2020,devora2024} but found no signal in the current dataset. We then tested the detectability with the MATRIX code by injecting and recovering synthetic planetary signals, using periods from 18 to 23 days and sizes from 1.5 to 5.5~\rearth\ \citep[see, e.g.,][]{devora2022,pozuelos2023}. As shown in Fig.~\ref{fig:TOI521_IR}, detectability depends strongly on the candidate's size. The nominal size from \texttt{spright} yields a $\sim$70\% recovery rate, increasing to 100\% for sizes above 3.5~\rearth. For sizes $\leq$ 3.0~\rearth, recovery drops below 10\%, indicating the candidate is undetectable in this range using the available dataset. Thus, we can currently only rule out a transiting body with sizes above 3.5~\rearth\ around this period.  

\section{Discussion}\label{sec:discussion}

\subsection{Mass, radius, and composition}\label{sec:thirstee_sample}

\begin{figure}[h!]
\centering
\includegraphics[width=0.9\linewidth]{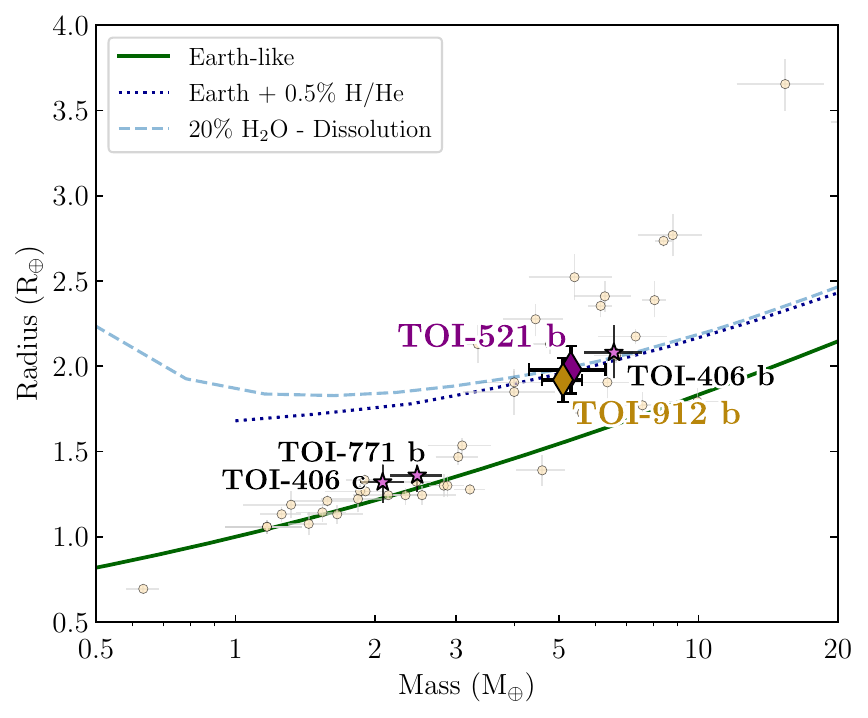}
  \caption{Mass-radius diagram for well-characterised sub-Neptunes around M dwarfs. Data are taken from the PlanetS catalogue \citep{parc2024}, including also more recently characterised planets from the NASA Exoplanet Archive (as of 17 July 2025), that have radii and masses measured with a precision better than 8\% and 25\%, respectively. Planets flagged as showing TTVs are removed from sample. The two planets characterized in this work are highlighted with purple and golden diamonds, and the additional planets of the \thirstee\ sample are reported with pink stars. The coloured lines represent different theoretical compositions: Earth-like composition (solid green line) from \citep{Zeng2019}; Earth-like cores with H/He envelopes at $F = 10$~F$_{\oplus
  }$ stellar flux (dotted dark blue line) from \cite{Lopez_2014}; a model including water dissolution in both core and mantle (dotted grey line) from \cite{Luo24}. 
  }
    \label{fig:MR_diagram}
\end{figure}

We show in Fig.~\ref{fig:MR_diagram} the position of the two transiting planets in the mass-radius diagram, comparing it with the known, well-characterised population of sub-Neptunes around M dwarfs. Considering the recent results of \cite{leleu24}, who report that (nearly) resonant sub-Neptunes, which is usually the case for TTV planets, are less dense than non-resonant systems, we removed from the plot all planets flagged as showing TTVs, as they may belong to a separate demographic population.

Interestingly, TOI-521 b and TOI-912 b , which orbit very similar stars (Sect.~\ref{sec:star}), have also almost identical mass and radius. 
They are located in the degenerate region of the mass-radius diagram where different internal structure and models can explain their composition. In fact, their planetary density could be both consistent with a significant amount of water in their interiors (up to $\sim 20$\%, according to \citep{Luo24} models including water dissolution in the planetary interiors), or with a rocky core plus a $\sim 0.5$\% H-He envelope \citep{Lopez_2014}. 
This degeneracy is further confirmed by the internal structure model we performed with the \texttt{ExoMDN}\footnote{\url{https://github.com/philippbaumeister/ExoMDN}} code \citep{Baumeister2023}, which computes through mixture density networks the mass and radius fraction of a fixed 4-layer planet (core, mantle, water, and atmosphere) given a planetary mass, radius and temperature. As Fig.~\ref{fig:ExoMDN} shows, both planets have high uncertainties on the water content, which could be up to $\sim 40-50$\% in mass fraction, and a negligible H/He envelope, which however can take up to $\sim 20-25$\% of the radius fraction. It is important to note that this computation has been performed assuming the equilibrium temperature of the planets reported in Table~\ref{table:joint_parameters}, which could not be accurate in the case of TOI-912 b due to tidal heating (see Sect.~\ref{sec:planet_properties_b}), which could imply a more extended envelope. 

\begin{figure}
\centering
  \includegraphics[width=\linewidth]{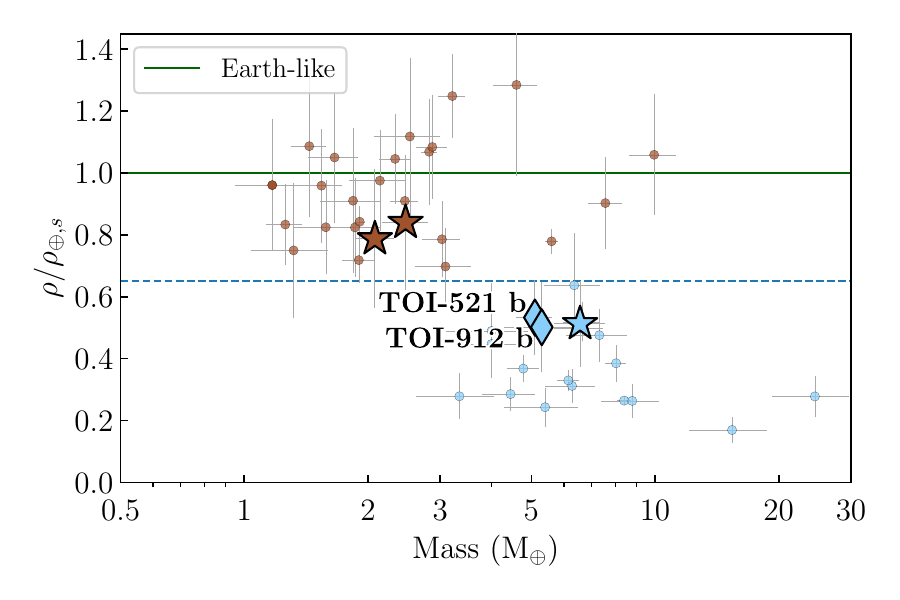}
  \caption{Mass-density of the M-dwarf population of sub-Neptunes, using the same sample as in Fig.~\ref{fig:MR_diagram}, adapted from \citep{luquepalle2022}. TOI-912 b and TOI-521 b are labelled and highlighted with diamonds, and the additional \thirstee\ planets are represented with stars. Planets are divided into two populations according to their density, assuming a density gap at $0.65$ $\rho_{\oplus, s}$ (dotted blue line), which according to \cite{luquepalle2022} divides the population of rocky planets (brown points) from a volatile (or gas-) enriched population (blue points). 
  }
    \label{fig:density_gap}
\end{figure}

While commonly the water-world hypothesis (i.e. \citealt{luquepalle2022}), and atmospheric mass-loss processes (i.e. \citealt{Rogers23}) have been used to explain the distribution and internal composition of this category of sub-Neptunes, more complex models that are currently being developed, including for example redox processes, hydrogen miscibility, magma ocean interactions, and core/mantle melting will be necessary to properly describe the physical processes happening in the interior and atmospheres of such planets (i.e. \citealt{DornLichtenberg21, Rogers2024, Luo24, gupta2025}). However, independently on the adopted internal model, when looking at the current \thirstee\ newly characterised exoplanets around M-dwarfs, which also includes aTOI-406 b, c \citep{lacedelli2024} and TOI-771 b \citep{lacedelli2025}, it seems that this sample supports the hypothesis formulated by \cite{luquepalle2022}, according to which sub-Neptunes are divided into at least two distinct density populations, with a density gap around $0.65$ $\rho_{\oplus, s}$\footnote{$\rho_{\oplus, s}$ is the scaled Earth bulk density, as defined in \cite{luquepalle2022}.} (Fig.~\ref{fig:density_gap}).

\begin{figure}
\centering
\includegraphics[width=\linewidth]{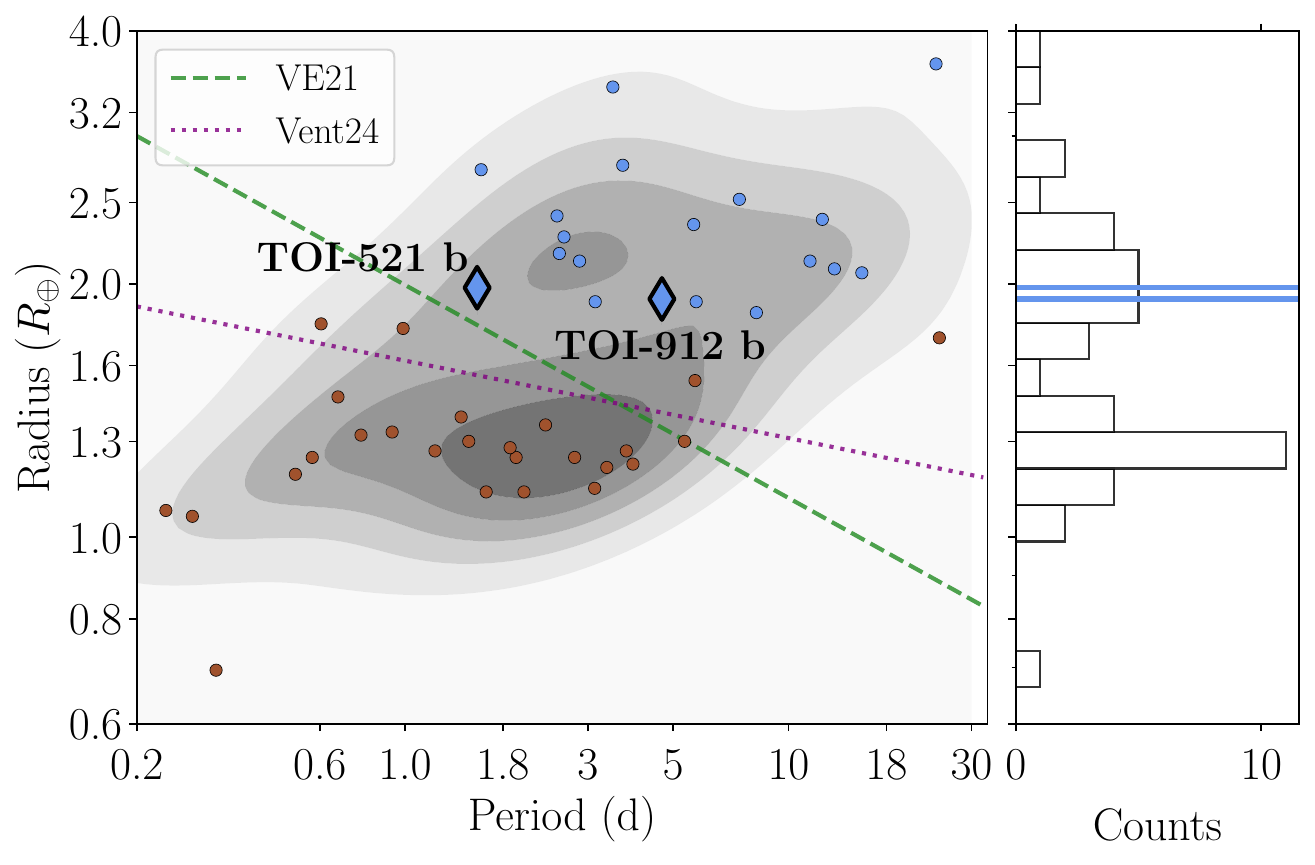}
\caption{Period-radius diagram of the M-dwarf population of sub-Neptunes, using the same sample as in Fig.~\ref{fig:MR_diagram}. Planets are colour-coded according to their density, as in Fig.~\ref{fig:density_gap}. The green-dashed and purple-dotted lines mark the position of the radius gap for M dwarfs estimated by \cite{VanEylen21} and \cite{venturini2024}, respectively. The two planets under analysis are highlighted with diamonds. The blue lines on the histogram, which shows the radius distribution of the sample, marks the radius values of the two targets.}\label{fig:period_radius}
\end{figure}

The division of the sub-Neptunes into different populations is also clear when looking at their radius distribution, with planets above the radius gap \citep{Fulton2017} being mostly volatile-enriched, and planets below the gap being consistent with a rocky composition (Fig.~\ref{fig:period_radius}). As expected, both TOI-521 b and TOI-912 b are located above the location of the radius gap for M dwarfs \citep{vanEylen2018, cloutier2020, venturini2024}. However, while TOI-912 b is located in a highly populated region of the diagram, TOI-521 b stands up as one of the shortest period sub-Neptunes above, but nearly in the radius gap according to the observational prediction of \cite{VanEylen21}. This makes it a very interesting target to study the evolution of planetary atmospheres at short orbital period, as well as for the modelling of the interior water mass fraction of highly irradiated planets (i.e. \citealt{aguichine2021}). 

\subsection{Eccentricity of TOI-912 b}\label{sec:planet_properties_b}
\begin{figure}[h!]
\centering
\includegraphics[width=0.9\linewidth]{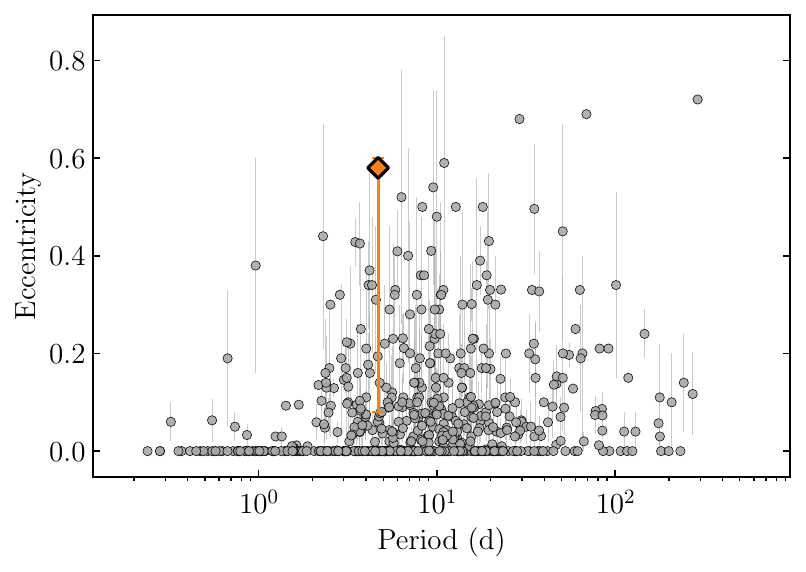}
  \caption{Eccentricity of all known exoplanets with \rplanet~$< 4$~\rearth\ (from the NASA Exoplanet Archive as of 17 July 2025) as a function of orbital period. TOI-912 b is highlighted with an orange diamond. The orange line marks a conservative lower limit to the eccentricity, encompassing all the values of the models discussed in Sect.~\ref{sec:rv_fit_only}.
  }
    \label{fig:period_ecc}
\end{figure}

While almost identical to TOI-521 b in mass and radius, TOI-912 b is very different in term of orbital properties. In fact, while for TOI-521 b our model suggests a circular orbit, TOI-912 b has an eccentricity of $e = 0.58 \pm 0.02$, which, if confirmed in the future (see Sect.~\ref{sec:rv_fit_only}), makes it one of the most eccentric sub-Neptunes discovered to date (Fig.~\ref{fig:period_ecc}). 
Recent discoveries on sub-Neptunes report the presence of similar short-period, eccentric planets, like TOI-5800 b  \citep{naponiello2025, jenkins2025}, which are thought to undergo high-eccentricity migration (HEM) moving into the Neptunian desert.
The circularization timescale $t_c$ \citep{goldreich1966, rasio1996} for TOI-912 b, assuming the derived stellar and planetary parameters and a modified tidal quality factor $Q'_p$ of $10^5$ (as the one estimated for Uranus and Neptune; \citealt{tittemore1990, banfield1992}), is around $\sim 14$ Gyr. 
This is consistent with the HEM scenario, according to which the planet migrated on a highly eccentric orbit, and it is still in the process of tidal circularization (i.e. \citealt{rasio1996}). An alternative explanation could be the presence of an additional, perturbing body in mean motion resonance, which can excite the eccentricity of planet b while maintaining a stable orbital configuration (i.e. \citealt{mardling2007}). However, according to our TTV analysis, there is no evidence of such a strongly interacting body in the current dataset. Nonetheless, considering the short observational baseline, additional observations will be essential to investigate the possibility of a currently undetected body. The high eccentricity could also be the result of giant impacts after disk dispersal \citep{shibata2025}. 
In this scenario, rocky planets (initially \rplanet$\lesssim 2$~\rearth) that undergo strong dynamical instabilities and numerous late giant impacts have their orbits excited (increased eccentricity) and their radii increased, ultimately placing them into the radius valley. This could explain the higher eccentricity (marginally) observed for planets within the radius valley that has been recently highlighted by \cite{gilbert2025}. With \rplanet $\sim 1.98$~\rearth, TOI-912 b seems to follow this trend.

Due to its high eccentricity, tidal dissipation plays an important role in the thermal structure of the planet. 
We estimated the tidal luminosity of the planet, following 
\cite{leconte2010} and assuming $Q'_p = 10^5$, to be $\sim 1.97 \times 10^{17}$W\footnote{This depends strongly on the assumed $Q'_p$, and it can be even higher considering that \cite{leconte2010} formulation is an approximation for low-eccentricity orbits.}.  
This is almost a few percent of the incident stellar flux received from the host star ($\sim 9.8 \times 10^{18}$ W), and it therefore implies an important additional internal heating source. This can significantly alter the thermal balance and internal/atmospheric properties of the planet, and especially the radius inflation in presence of an atmosphere.  

Given the results of our RV analysis in Sect.~\ref{sec:rv_fit_only}, we stress that additional, well-sampled observations will be needed to definitely confirm the eccentricity of the planet.

\subsection{Prospects for atmospheric characterisation}\label{sec:atmosphere}
We computed the transmission and emission spectroscopy metrics (TSM, ESM;  \citealt{kempton2018}) to investigate the suitability of the targets for atmospheric characterization. With TSM = 83.5 and ESM = 5.4, TOI-912 b is at the lowest limit of the observability thresholds\footnote{\citet{kempton2018} propose ESM > 7.5, and TSM > 90 for $ 1.5 < $~\rearth\ \rplanet $ < 2.75 $~\rearth\ for optimal target selection.}, making it a suitable candidate for both transmission and emission studies, even though not an optimal one. However, it is worth noting that both metrics are computed assuming the equilibrium temperature of the planet based only on stellar irradiation, while the planet is likely to have a much higher temperature due to tidal dissipation (see Sect.~\ref{sec:planet_properties_b}), making it a very interesting target also to study atmospheric mass loss processes. 

Similarly, TOI-521 b has TSM = 54.1 and ESM = 6.9, making it a suitable target for \textit{JWST} studies, especially in emission spectroscopy.

Even though slightly more massive, TOI-521 b and TOI-912 b have similar properties to GJ 9827 d (\teq $\sim 618$~K, \rplanet $=1.98$~\rearth, \mplanet $= 3.02$~\mearth), for which atmospheric features have been recently identified with \textit{JWST} \citep{Piaulet2024}. These planets will therefore be interesting targets to test the compositional trend with
temperature suggested by \cite{Piaulet2024}, hinting to a decrease in the H$_2$-He feature strength from colder to warmer sub-Neptunes.
\section{Conclusions}\label{sec:conclusions}
In this work, we characterized two new sub-Neptunes, TOI-521 b and TOI-912 b, as part of the \texttt{THIRSTEE} sample, using \textit{TESS} data,  ground-based photometry, and high-precision ESPRESSO, HARPS and IRD RVs. 
Both planets orbit similar M3 dwarfs, and have almost identical radii ($1.98 \pm 0.14$~\rearth; $1.93 \pm 0.13$~\rearth) and masses ($5.3 \pm 1.0$~\mearth; $5.1 \pm 0.5$~\mearth), implying a median bulk density of $\sim 4$ \gcm. With such a density, they lie within a highly degenerate region of the mass-radius diagram, where both a volatile-rich and a H/He envelope composition can explain their bulk properties.
Even though both planets have short orbital periods ($P = 1.5$~d and $P= 4.7$~d, respectively), TOI-521 b shows a circular orbit and it is located very close to the radius gap line, while TOI-912 b is (likely) one of the most eccentric ($e = 0.58 \pm 0.02$) sub-Neptunes discovered up to date, making it a very interesting target for tidal interaction studies.
Finally, we identified a second candidate in a 20.3-d orbit in the TOI-521 system, with a minimum mass of $M_{\rm p} \sin{i} = 10.7$~\mearth, placing it between the sub-Neptune and Neptune regime, even though currently not detected in transit. Given the low-significance of the models comparison, more observations are needed to confirm it as a planet. 

Both TOI-521 b and TOI-912 b add up to the sample of well-characterized sub-Neptunes around M-dwarfs, probing the effectiveness of the \texttt{THIRSTEE} program, and they further strengthen the hypothesis of a density gap among the M-dwarf population \citep{luquepalle2022, Schulze2024}, as the previous \texttt{THIRSTEE} targets. Finally, these sibling planets are warm sub-Neptunes (\teq$ = 794$~K and $551$ K), and they are both suitable for atmospheric characterisation studies with \textit{JWST}, especially in the context of investigating the compositional trends with temperature among the sub-Neptune population.

\section*{Data availability}\label{sec:data_availability}
The ESPRESSO, HARPS and IRD RVs and activities indicators are available in electronic form at the CDS via anonymous ftp to cdsarc.u-strasbg.fr (130.79.128.5) or via \url{http://cdsweb.u-strasbg.fr/cgi-bin/qcat?J/A+A/}.

\begin{acknowledgements}
We thank the referee for the comments which improved the quality of this manuscript.
We acknowledge financial support from the Agencia Estatal de Investigaci\'on of the Ministerio de Ciencia e Innovaci\'on MCIN/AEI/10.13039/501100011033 and the ERDF “A way of making Europe” through project PID2021-125627OB-C32, and from the Centre of Excellence “Severo Ochoa” award to the Instituto de Astrofisica de Canarias. We acknowledge funding from the European Research Council under the ERC Grant Agreement n. 337591-ExTrA.
This paper includes data collected by the {\it TESS} mission,
which are publicly available from the Mikulski Archive for Space
Telescopes (MAST). Funding for the {\it TESS} mission is provided 
by the NASA Explorer Program. 
Resources supporting this work were provided by the NASA High-End Computing (HEC) Program through the NASA Advanced Supercomputing (NAS) Division at Ames Research Center for the production of the SPOC data products. We acknowledge the use of public TESS data from pipelines at the TESS Science Office and at the TESS Science Processing Operations Center.
This research has made use of the Exoplanet Follow-up Observation Program (ExoFOP; DOI: 10.26134/ExoFOP5) website, which is operated by the California Institute of Technology, under contract with the National Aeronautics and Space Administration under the Exoplanet Exploration Program. 
Funding for the TESS mission is provided by NASA's Science Mission Directorate. 
This research has made extensive use of the SIMBAD
database, operated at CDS, Strasbourg, France, and NASA’s Astrophysics Data System.
This research has
made use of the NASA Exoplanet Archive, which is
operated by the California Institute of Technology, 
under contract with the National Aeronautics and Space
Administration under the Exoplanet Exploration Program. 
This work has made use of data from the European Space Agency (ESA) mission
{\it Gaia} (\url{https://www.cosmos.esa.int/gaia}), processed by the {\it Gaia}
Data Processing and Analysis Consortium (DPAC,
\url{https://www.cosmos.esa.int/web/gaia/dpac/consortium}). 
Funding for the DPAC
has been provided by national institutions, in particular the institutions
participating in the {\it Gaia} Multilateral Agreement.
This publication makes use of data products from the Two
Micron All Sky Survey, which is a joint project of the
University of Massachusetts and the Infrared Processing and
Analysis Center/California Institute of Technology, funded by
the National Aeronautics and Space Administration and the
National Science Foundation. 
This work makes use of observations from the LCOGT network. Part of the LCOGT telescope time was granted by NOIRLab through the Mid-Scale Innovations Program (MSIP). MSIP is funded by NSF.
Some of the observations in this paper made use of the High-Resolution Imaging instrument Zorro and were obtained under Gemini LLP Proposal Number: GN/S-2021A-LP-105. Zorro was funded by the NASA Exoplanet Exploration Program and built at the NASA Ames Research Center by Steve B. Howell, Nic Scott, Elliott P. Horch, and Emmett Quigley. Zorro was mounted on the Gemini South telescope of the international Gemini Observatory, a program of NSF's OIR Lab, which is managed by the Association of Universities for Research in Astronomy (AURA) under a cooperative agreement with the National Science Foundation. on behalf of the Gemini partnership: the National Science Foundation (United States), National Research Council (Canada), Agencia Nacional de Investigación y Desarrollo (Chile), Ministerio de Ciencia, Tecnología e Innovación (Argentina), Ministério da Ciência, Tecnologia, Inovações e Comunicações (Brazil), and Korea Astronomy and Space Science Institute (Republic of Korea).
This research is based on data collected at the Subaru Telescope, which is operated by the National Astronomical Observatory of Japan. We are honored and grateful for the opportunity of observing the Universe from Maunakea, which has the cultural, historical and natural significance in Hawaii.
This paper is based on observations with the MuSCAT2 instrument, developed by Astrobiology Center, at Telescopio Carlos S\'{a}nchez operated on the island of Tenerife by the IAC in the Spanish Observatorio del Teide.
This study was partly supported by the JSPS KAKENHI Grant Numbers JP19KK0082, JP20K14521, JP21K13955, JP21K13987, JP23H00133, JP23H01224, JP23H01227, JP23K17709, JP23K25920, JP23K25923, JP24H00017, JP24H00242, JP24H00248, JP24K00689, JP24K07108, JP24K17082, JP24K17083, JP25H00005, JP25K01061, JP25K17450, JSPS Bilateral Program Number JPJSBP120249910, JSPS Grant-in-Aid for JSPS Fellows Grant Number JP24KJ0241, JP25KJ0091, JP25KJ1036, JP25KJ1040, JST SPRING Grant Number JPMJSP2108.
The MEarth Team gratefully acknowledges funding from the David and Lucile Packard Fellowship for Science and Engineering (awarded to D.C.). This material is based upon work supported by NSF under grant AST-1616624, and by NASA under grant No. 80NSSC18K0476 (XRP Program). This work is made possible by a grant from the John Templeton Foundation. The opinions expressed in this publication are those of the authors and do not necessarily reflect the views of the John Templeton Foundation.

NASA NHFP — Support for this work was provided by NASA through the NASA Hubble Fellowship grant HST-HF2-51559.001-A awarded by the Space Telescope Science Institute, which is operated by the Association of Universities for Research in Astronomy, Inc., for NASA, under contract NAS5-26555. 
R.L. acknowledges financial support from the Severo Ochoa grant CEX2021-001131-S funded by MCIN/AEI/10.13039/501100011033. R.L. is funded by the European Union (ERC, THIRSTEE, 101164189). Views and opinions expressed are however those of the author(s) only and do not necessarily reflect those of the European Union or the European Research Council. Neither the European Union nor the granting authority can be held responsible for them.
D.J., K.G. and G.N. gratefully acknowledge the Centre of Informatics  Tricity Academic Supercomputer and networK (CI TASK, Gda\'nsk, Poland)  for computing resources (grant no. PT01187).
The work of HPO has been carried out within the framework of the NCCR PlanetS supported by the Swiss National Science Foundation under grants 51NF40\_182901 and 51NF40\_205606.
N.A-D. acknowledges the support of FONDECYT project 1240916.

\end{acknowledgements}

%
%

\bibliographystyle{aa}
\bibliography{bibliography}

\begin{appendix} 

\section{Photometric data}\label{appendix:light_curves}
\begin{figure}[h!]
\centering
\includegraphics[width=0.9\linewidth]{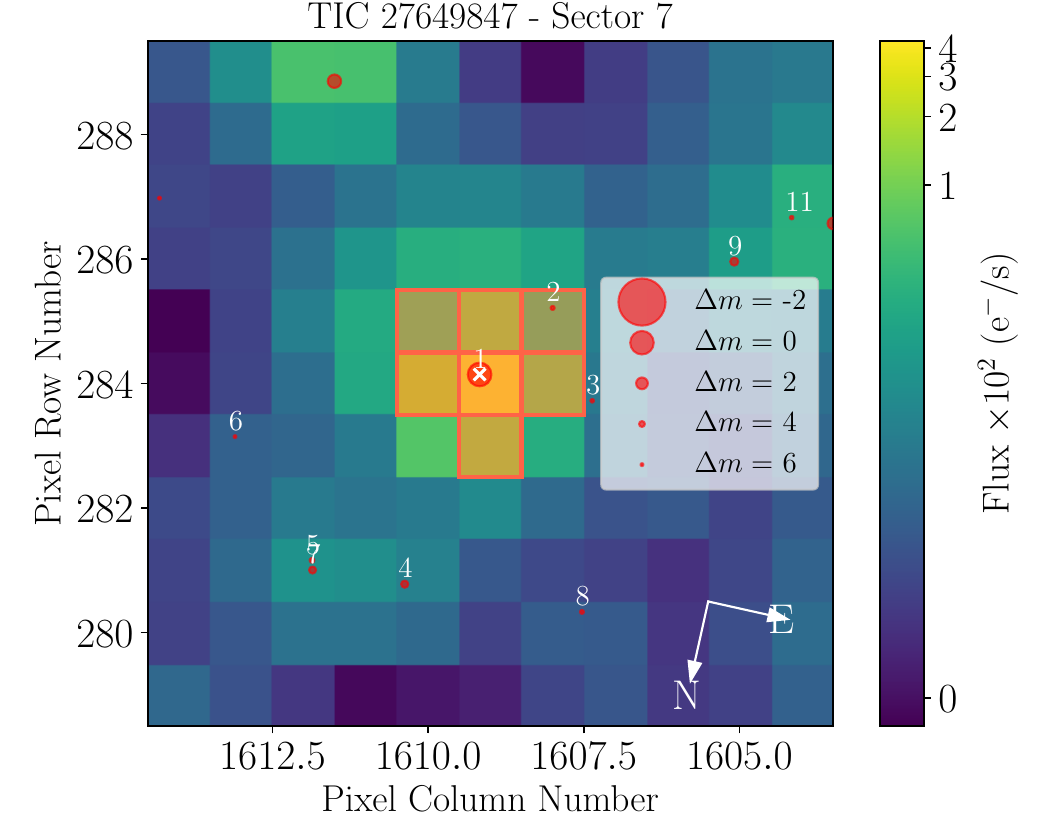}
\includegraphics[width=0.9\linewidth]{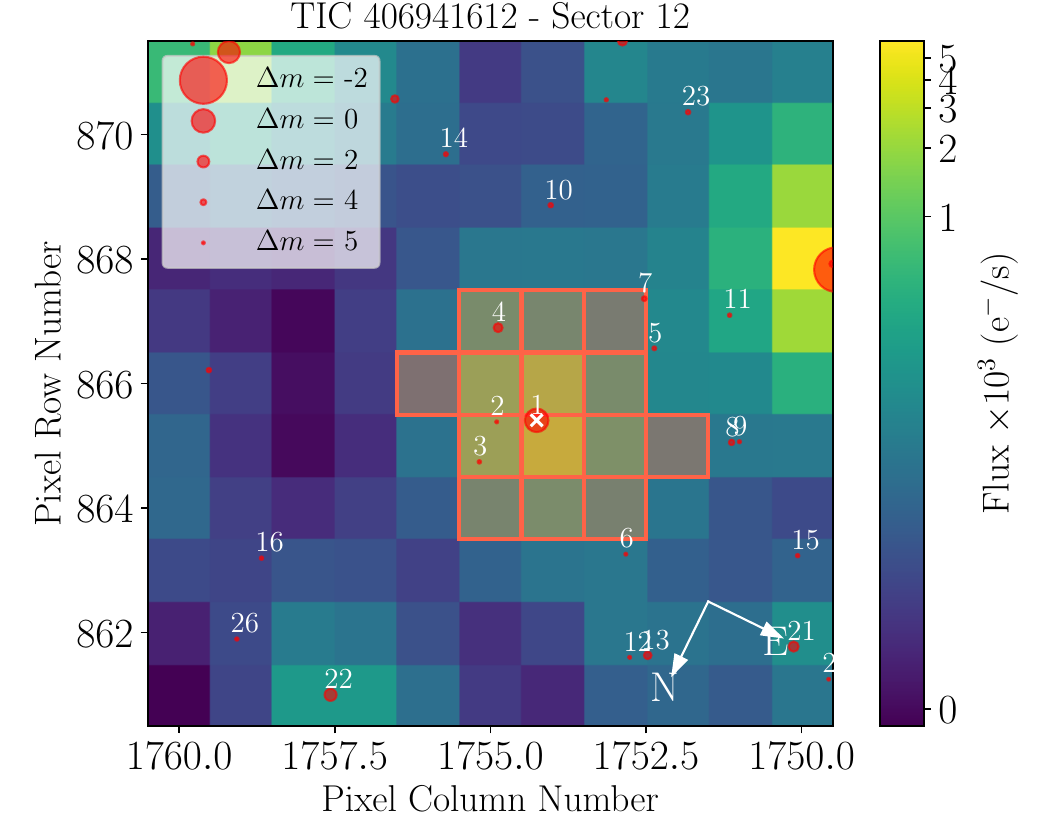}
\caption{TPF images of TOI-521 (TIC 27649847, sector 7) and TOI-912 (TIC 406941612, sector 12). In each image, the target star is labelled as 1, and all {\it Gaia} DR3 \citep{GaiaColl2023} sources in the field of view with magnitude contrast up to $\Delta m = 6$ are shown with red circles. The photometric aperture used for the light curve extraction is highlighted with red squares. Figures are produced with \texttt{tpfplotter} \citep{Aller2020}.}\label{fig:tpf}
\end{figure}

We report here additional analysis and figures related to the photometric data collected for TOI-521 and TOI-912. 
Figure~\ref{fig:tpf} shows the \textit{TESS} field of view of each target, with the photometric aperture used for data extraction.

\begin{figure}
\centering
  \includegraphics[width=0.85\linewidth]{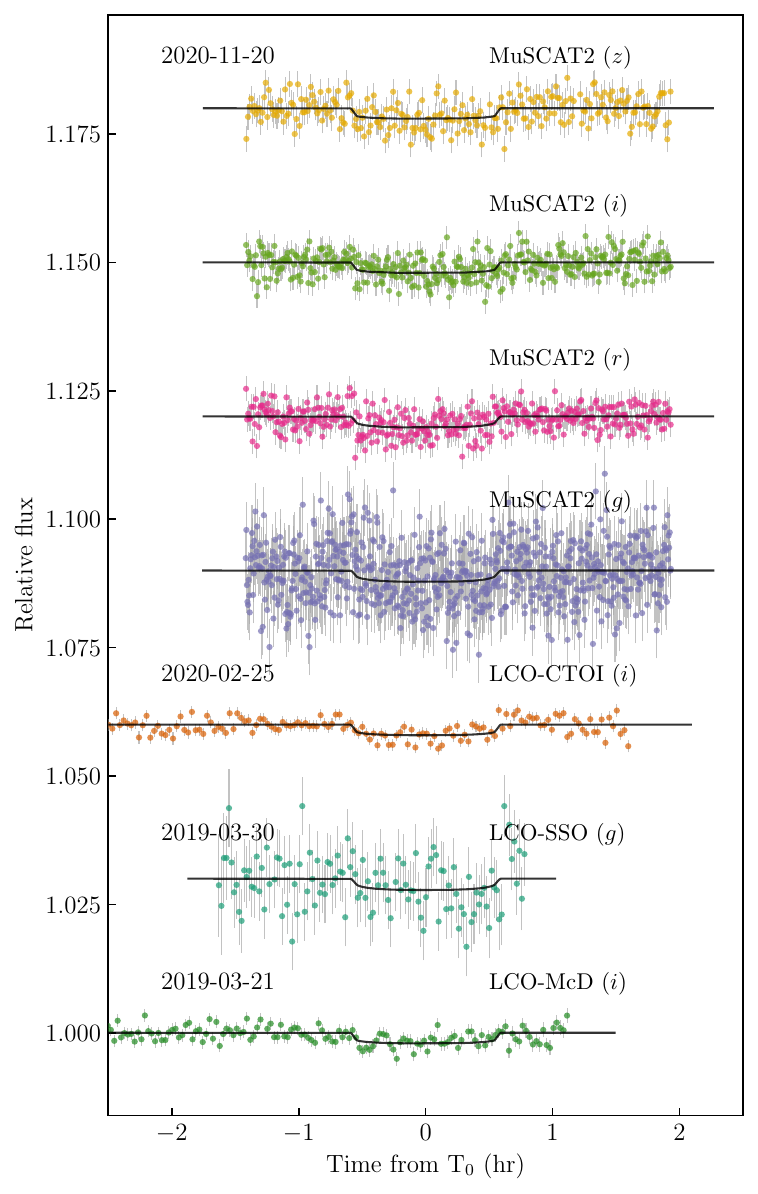}
  \includegraphics[width=0.85\linewidth]{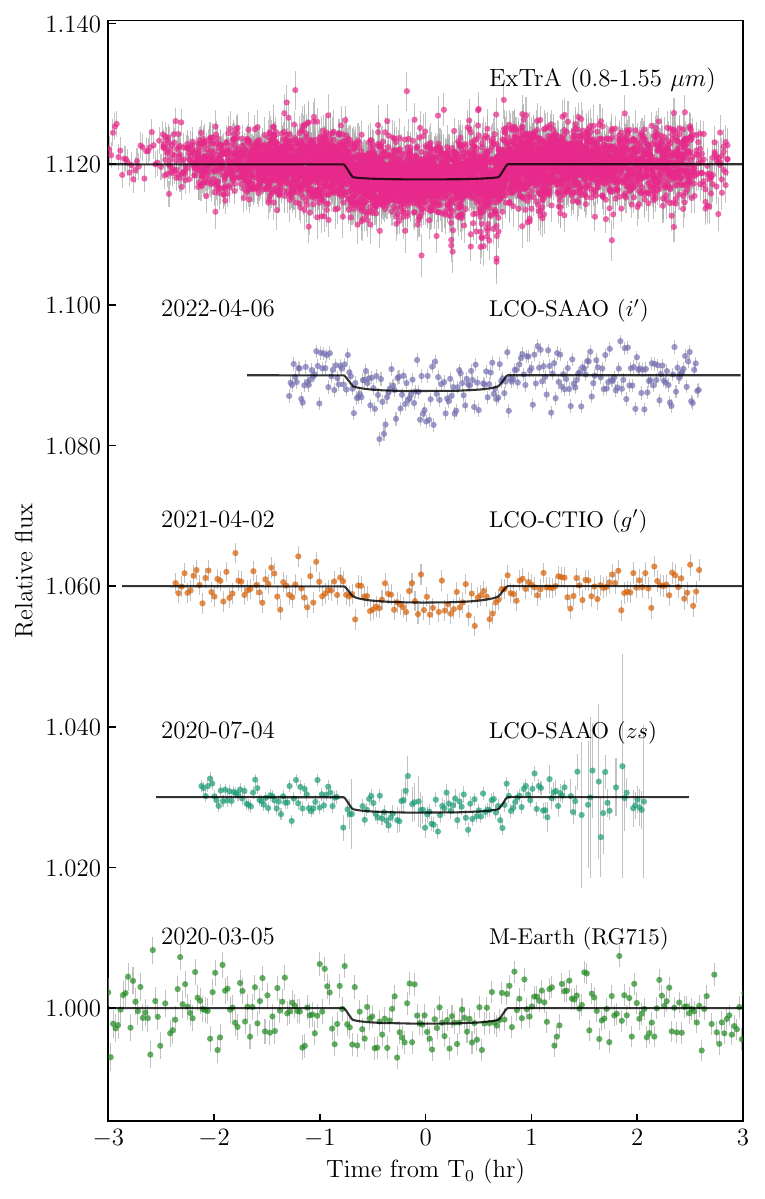}
  \caption{Ground-based photometry of TOI-521 b (top) and TOI-912 b (bottom). The best-fitting model is plotted as a solid black line. 
  }
    \label{fig:ground_based}
\end{figure}

Table~\ref{table:photometric_observations} summarizes the photometric transit data of the planets under analysis, Fig.~\ref{fig:ground_based} shows the results of the ground-based modelling performed in Sect.~\ref{sec:joint_fit}.

\begin{table}
\caption{Summary of the photometric observations for TOI-521 and TOI-912}
\centering
\begin{tabular}{lll}
\hline\hline
Instrument$^a$ & Date (UTC) & Filter  \\
\hline
TOI-521 b\\
\hline
\tess\ (Sector 7) & 2019 Jan-Feb & $T$ \\
\tess\ (Sector 34) & 2021 Jan-Feb & $T$ \\
\tess\ (Sector 44, 45, 46) & 2021 Oct-Dec & $T$ \\
\tess\ (Sector 71, 72) & 2023 Oct-Dec & $T$ \\
LCO-McD 1.0\,m   & 2019 Mar 21   & Sloan $i'$  \\
LCO-SSO 1.0\,m   & 2019 Mar 30   & Sloan $g'$    \\
LCO-CTIO 1.0\,m  & 2020 Feb 25   & ${I}$  \\
TCS-Muscat2 1.52\,m & 2020 Nov 20   & g, r, i, z  \\[2mm] 
\hline
TOI-912 b\\
\hline
\tess\ (Sector 12, 13)  & 2019 May-July & $T$ \\
\tess\ (Sector 39) & 2021 May-June & $T$ \\
\tess\ (Sector 65, 66) & 2023 May-July & $T$ \\
\tess\ (Sector 93) & 2025 June & $T$ \\
MEarth South 0.6\,m & 2020 Mar 05 & RG715 \\ 
LCO-SAAO 1.0\,m & 2020 July 04   & Pan-STARRS $z_s$ \\
LCO-CTIO 1.0\,m & 2021 Apr 21   & Sloan $g'$  \\
LCO-SAAO 1.0\,m & 2021 Apr 06   & Sloan $i'$   \\
ExTrA 0.6\,m (2 tel.) & 2021 Apr 29  & 0.8–1.55 $\mu$m \\
ExTrA 0.6\,m (2 tel.) & 2021 May 13  & 0.8–1.55 $\mu$m  \\
ExTrA 0.6\,m (1 tel.) & 2021 July 13 &  0.8–1.55 $\mu$m \\
ExTrA 0.6\,m (2 tel.) & 2022 June 10 &  0.8–1.55 $\mu$m \\ 
ExTrA 0.6\,m (2 tel.) & 2023 Mar 27 &  0.8–1.55 $\mu$m \\
ExTrA 0.6\,m (2 tel.) & 2023 Apr 10 &  0.8–1.55 $\mu$m \\
ExTrA 0.6\,m (3 tel.) & 2023 Apr 24 &  0.8–1.55 $\mu$m \\ 
ExTrA 0.6\,m (3 tel.) & 2023 June 10 &  0.8–1.55 $\mu$m \\   
ExTrA 0.6\,m (2 tel.) & 2023 June 24 &  0.8–1.55 $\mu$m \\
ExTrA 0.6\,m (3 tel.) & 2023 Aug 24 &  0.8–1.55 $\mu$m \\[2mm]
\hline
\end{tabular}
\tablefoot{
    \tablefoottext{a}{For the \tess\ data we report the observed sectors. For the ExTrA data we list the number of observing telescopes.} 
    }
\label{table:photometric_observations}
\end{table}

\begin{figure}
\centering
  \includegraphics[width=\linewidth]{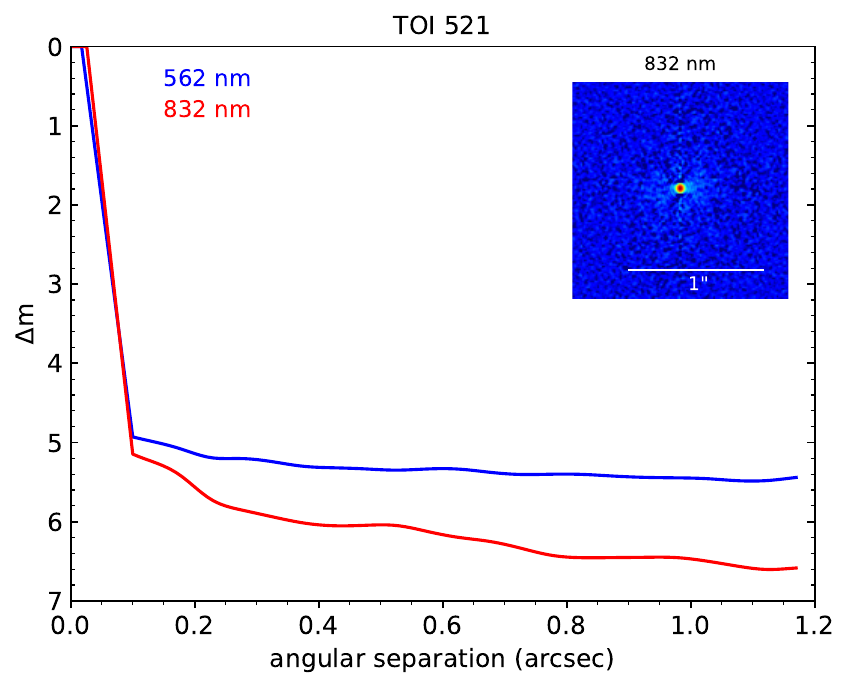}
  \includegraphics[width=\linewidth]{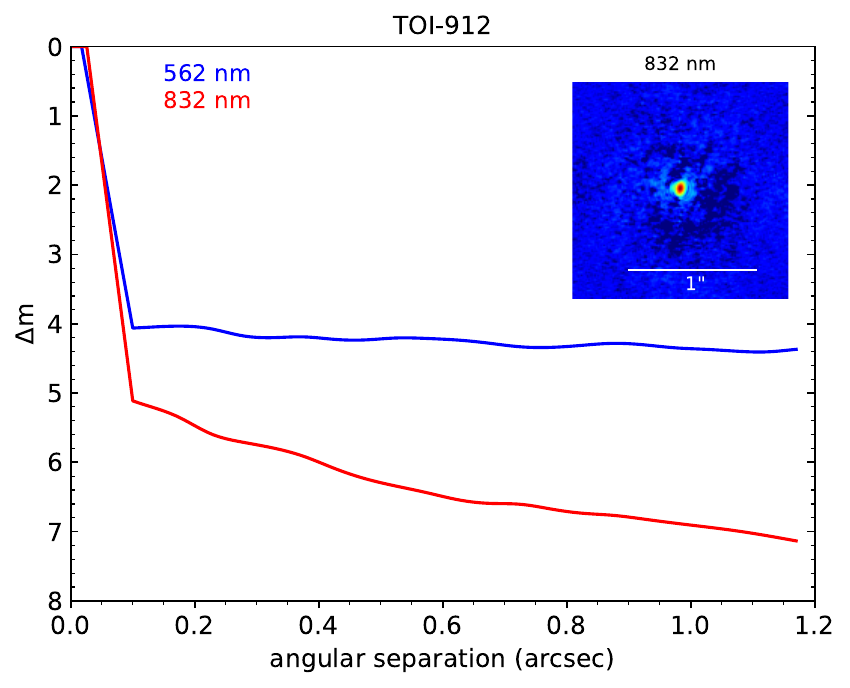}
  \caption{5$\sigma$ magnitude contrast curves of the two stars in both filters as a function of the angular separation out to 1.2 arcsec. In each plot, the inset shows the reconstructed 832 nm image with a 1-arcsec scale bar.
  }
    \label{fig:speckle}
\end{figure}

Additionally, we report the high-resolution speckle imaging we obtained  to search for confounding close companions to both stars. 
The presence of such a companion, whether truly bound or line of sight, provides ``third-light" contamination of the observed transit, leading to incorrect derived properties for the exoplanet and its host star 
\citep{Ciardi_2015, Furlan_2017, Furlan_2020}. 
The stars were observed on 09 December, 2021 (TOI-521) and March 14, 2020 (TOI-912) using the identical speckle instruments, 'Alopeke and Zorro respectively, mounted on the on the Gemini North/South 8-m telescopes \citep{scott2021}.
Both instruments provide simultaneous speckle imaging in two bands (562 nm and 832 nm were used here) with output data products that include robust 5$\sigma$ magnitude contrast curves and a reconstructed image.
Five sets of 1000 $\times$ 0.06 s images were obtained for each star. 
Data were processed with the standard reduction pipeline \citep{Howell2011} and Fig.~\ref{fig:speckle} shows the final 5$\sigma$ magnitude contrast curves and the 832 nm reconstructed speckle images for both stars. 
We find that TOI-521 and TOI-912 are both single stars with no companion brighter than 5-6.5 magnitudes and 5-7 magnitudes, respectively, below that of the target star from the Gemini Telescope 8-m telescope diffraction limit (20 mas) out to 1.2 arcsec. At the distance of TOI-521 ($d=$61 pc) and TOI-912 ($d =$26 pc), these angular limits correspond to spatial limits of 1.2-73 AU and 0.5-31 AU, respectively.

\section{Stellar analysis}\label{sec:stellar_abundance}
We report here some additional results from our stellar analysis.

\begin{figure}
\centering
  \includegraphics[width=\linewidth]{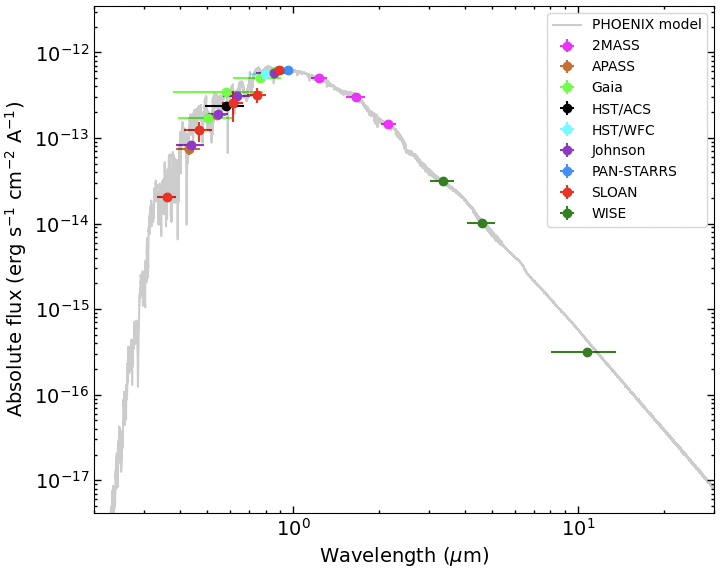}
  \includegraphics[width=\linewidth]{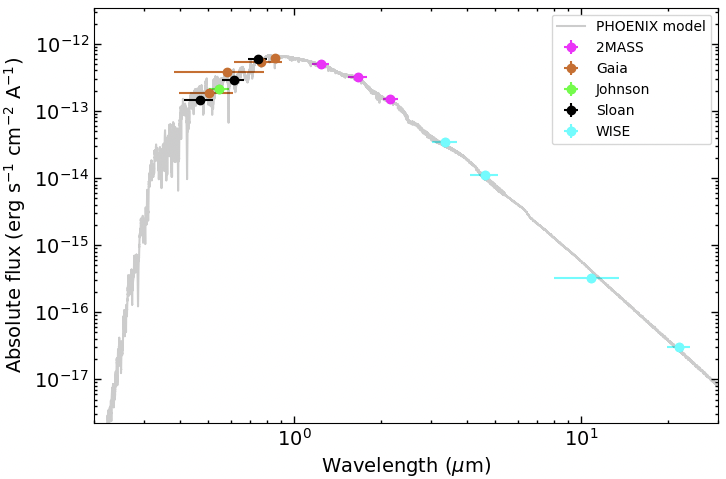}
  \caption{SED of TOI-521 (top) and TOI-912 (bottom). In both panels, the BT-Settl model computed assuming \teff~$=3500$~K, \logg~$= 5.0$, and solar metallicity is shown with a grey line. 
  }
    \label{fig:sed}
\end{figure}

\begin{figure}[h!]
\centering
  \includegraphics[width=\linewidth]{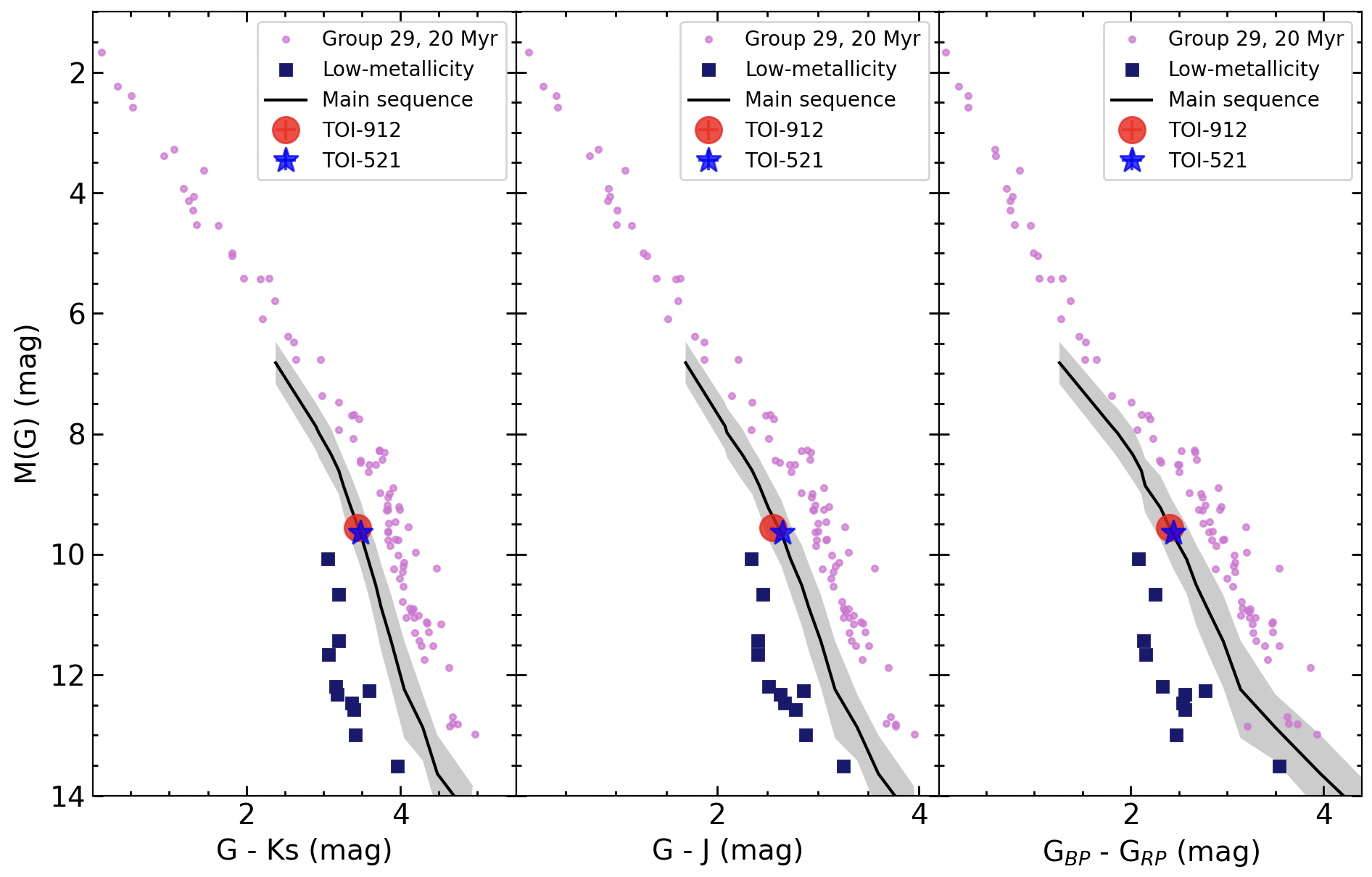}
  \caption{Color-magnitude diagrams of TOI-521 and TOI-912. The location of main-sequence M-type stars is indicated by the black line, while the gray area accounts for the observed dispersion \citep{Cifuentes20}. Young stars of the stellar moving Group 29 \citep{Luhman18} and old low-metallicity stars of similar colors \citep{Leggett00} are also shown for comparison purposes. TOI-521 and TOI-912 appear to be normal main-sequence stars.
  }\label{fig:m_stars}
\end{figure}

Figure~\ref{fig:sed} shows the SED fitting of TOI-521 and TOI-912 as described in Sect.~\ref{sec:star_properties}. Figure~\ref{fig:m_stars} shows the position of the two stars in the colour-magnitude diagram of M-dwarfs. 

\section{Periodogram analysis}\label{sec:periodograms}

\begin{figure*}
\centering
  \includegraphics[width=\linewidth]{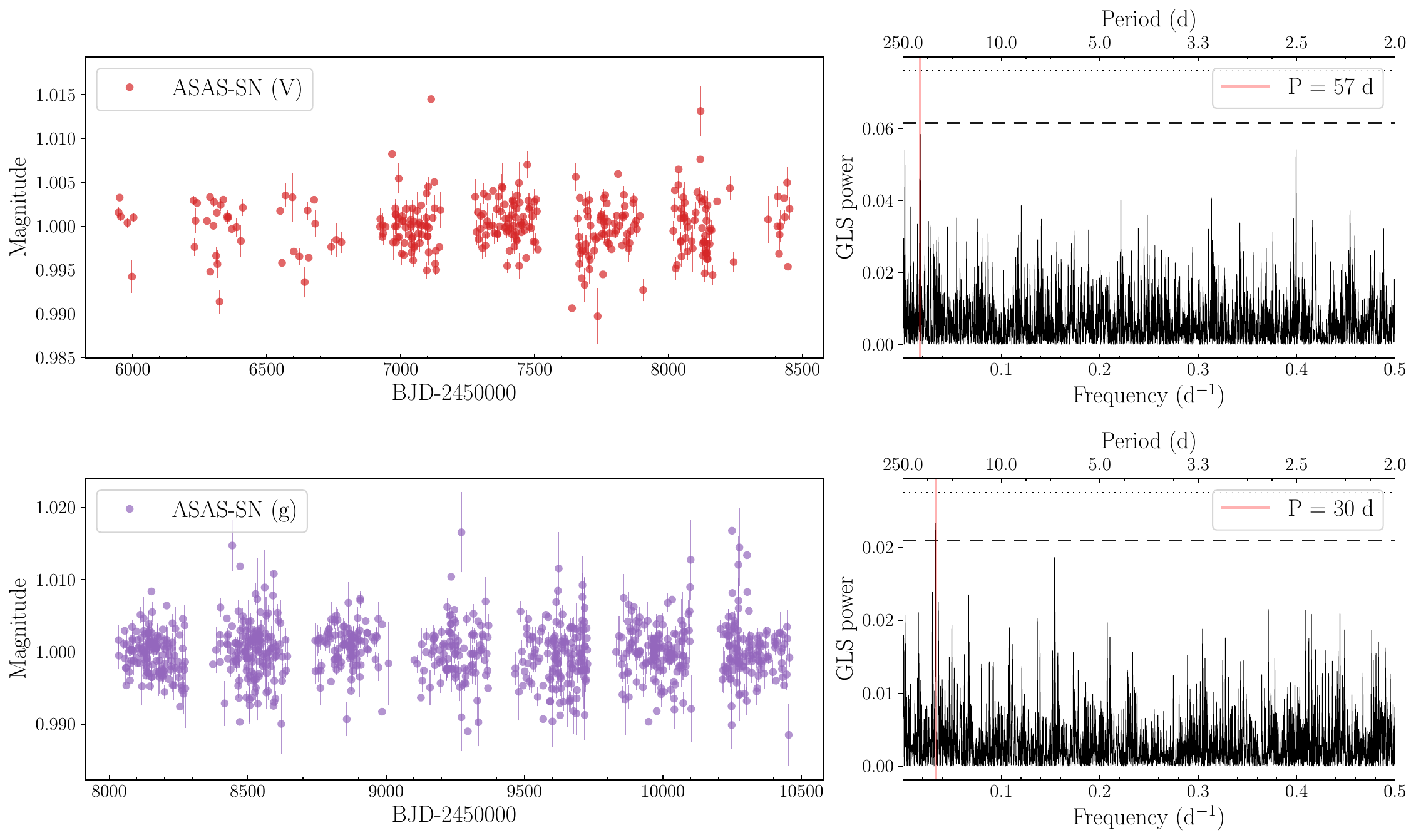}
  \includegraphics[width=\linewidth]{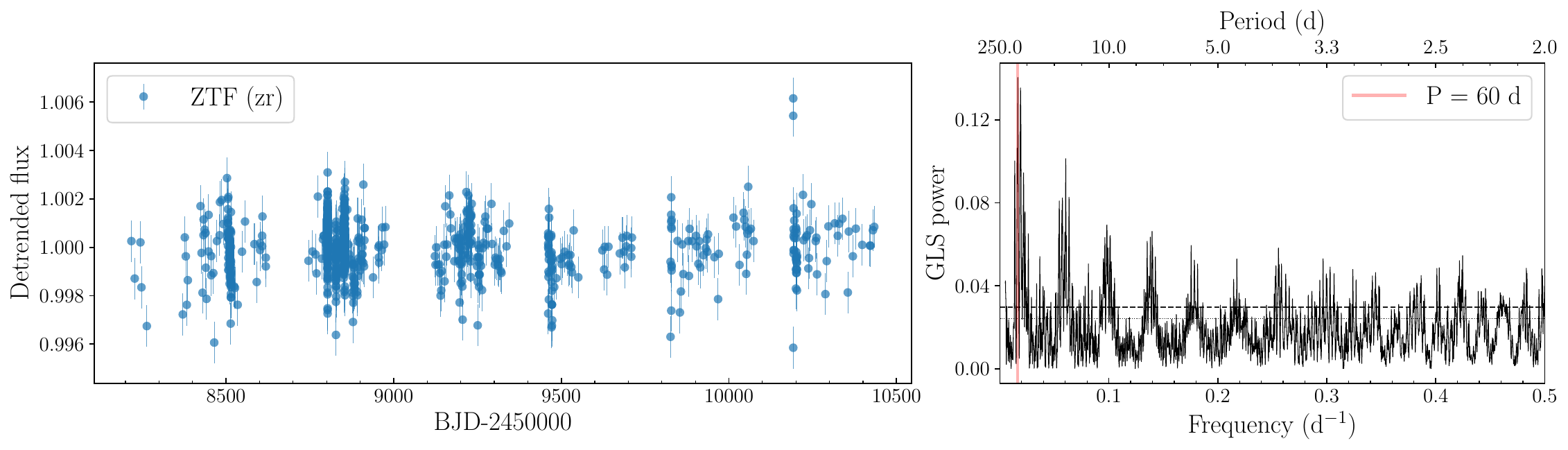}
  \caption{ASAS-SN (top and middle) and ZTF photometry (bottom) of TOI-521. The light curves are linearly detrended to remove the effect of the high proper motions of the star. The right panel of each plot shows the GLS periodogram, highlighting the $1$\% (dashed line) and $10$\% (dotted line) FAP levels. The most significant frequency is marked with a red vertical line. None of the ASAS-SN peaks is highly significant, while for ZTF many peaks and aliases are identified, probably due to the irregular sampling. 
  }\label{fig:asas_sn_toi521}
\end{figure*}

\begin{figure*}
\centering
\includegraphics[width=\linewidth]{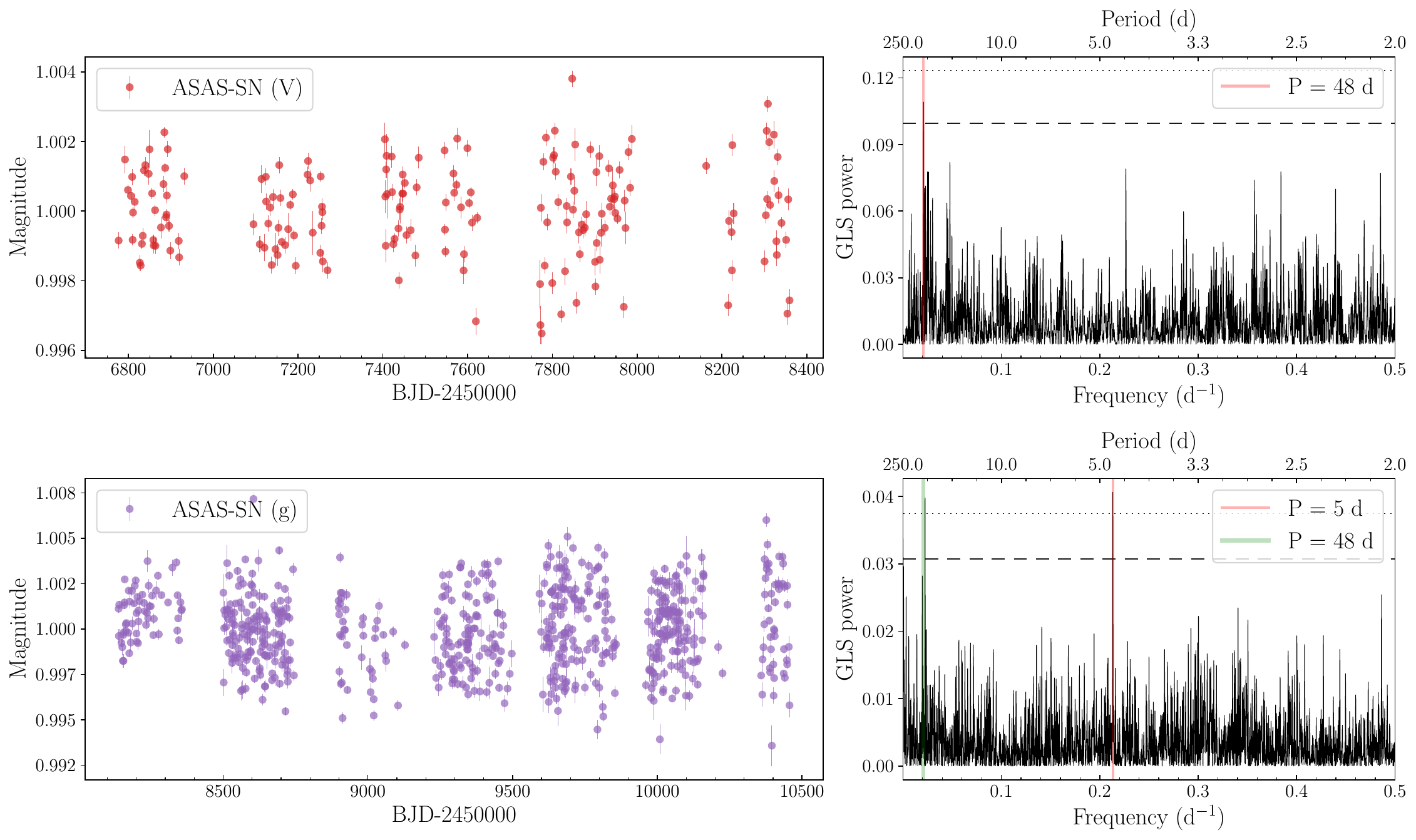}
  \includegraphics[width=\linewidth]{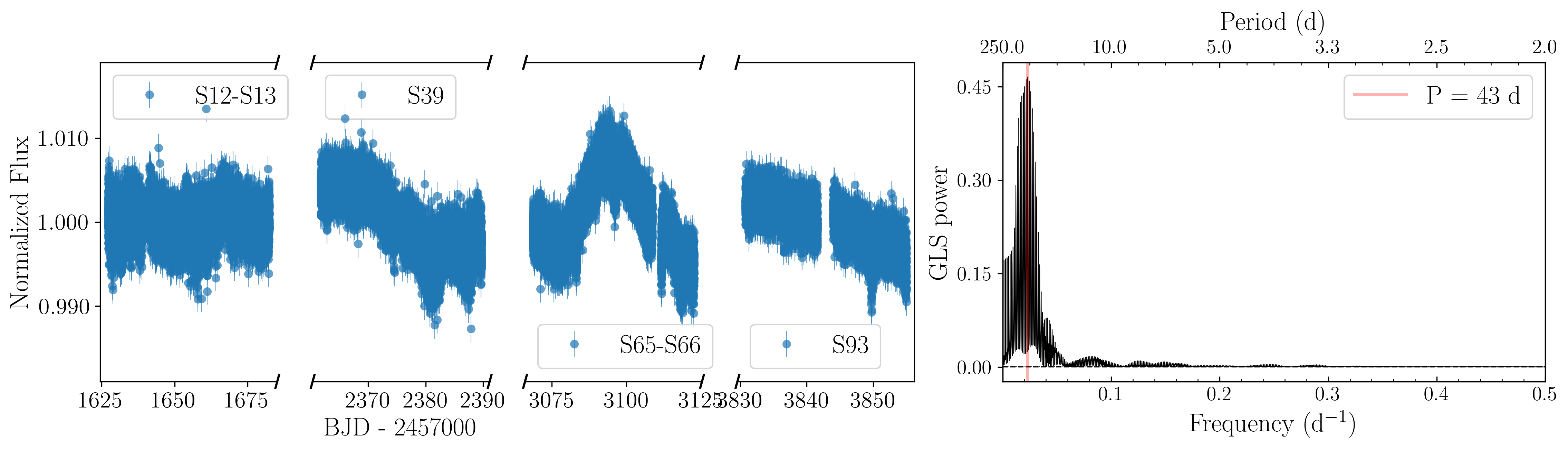}
  \caption{Same as Fig.\ref{fig:asas_sn_toi521}, but for the ASAS-SN (top and middle)} and \textit{TESS} SAP (bottom) photometry of TOI-912. The vertical green line shows the possible rotational period of the star.
  \label{fig:asas_sn_TOI912}
\end{figure*}

\begin{figure}[h!]
\centering
  \includegraphics[width=\linewidth]{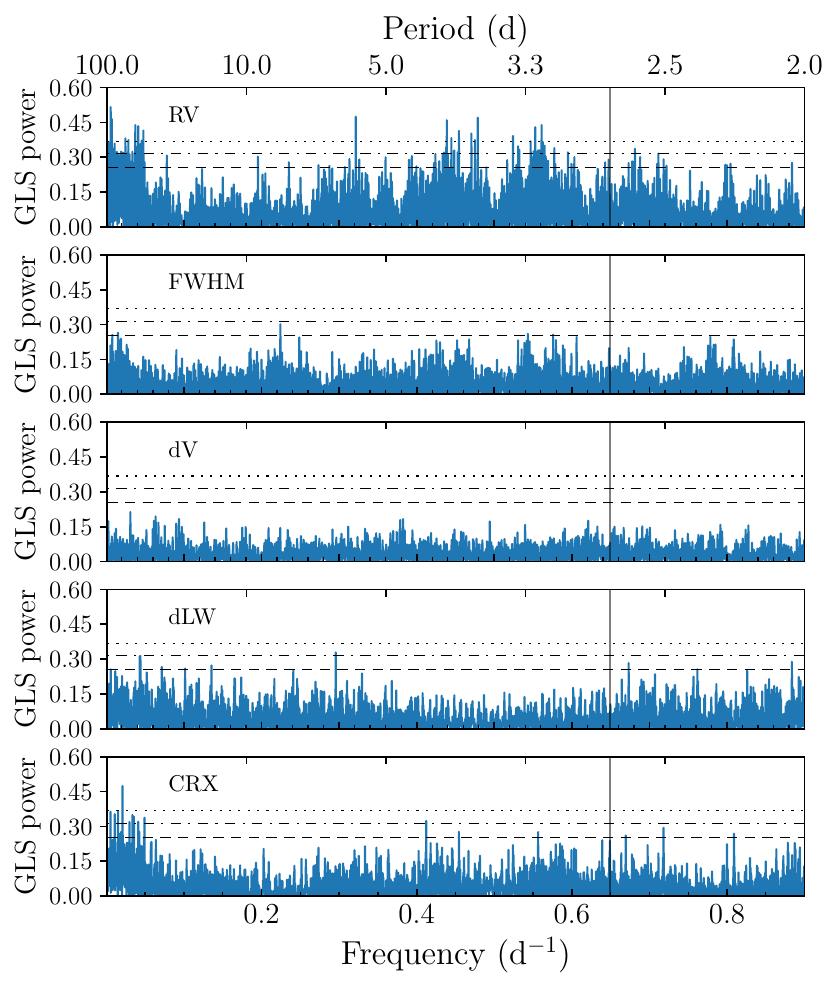}
  \includegraphics[width=\linewidth]{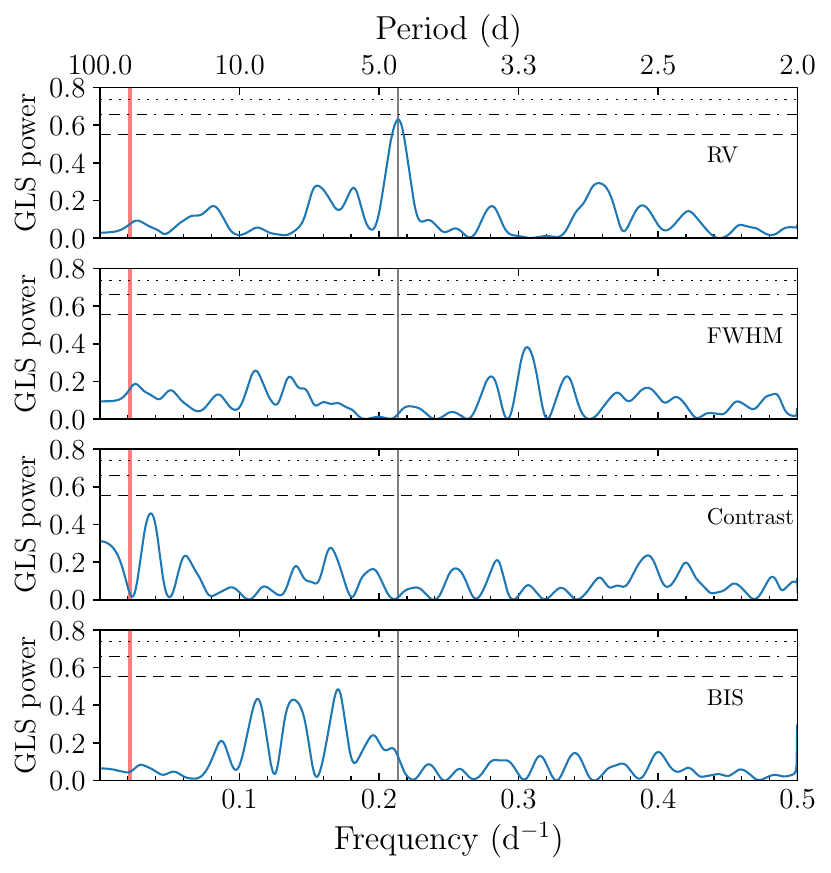}
  \caption{Same as Fig. \ref{fig:TOI912_periodogram}, but for the IRD dataset of TOI-521 (top) and the HARPS dataset of TOI-912 (bottom). 
  }
\label{fig:TOI521_IRD_HARPS_periodogram}
\end{figure}

\begin{figure}[h!]
\centering
  \includegraphics[width=\linewidth]{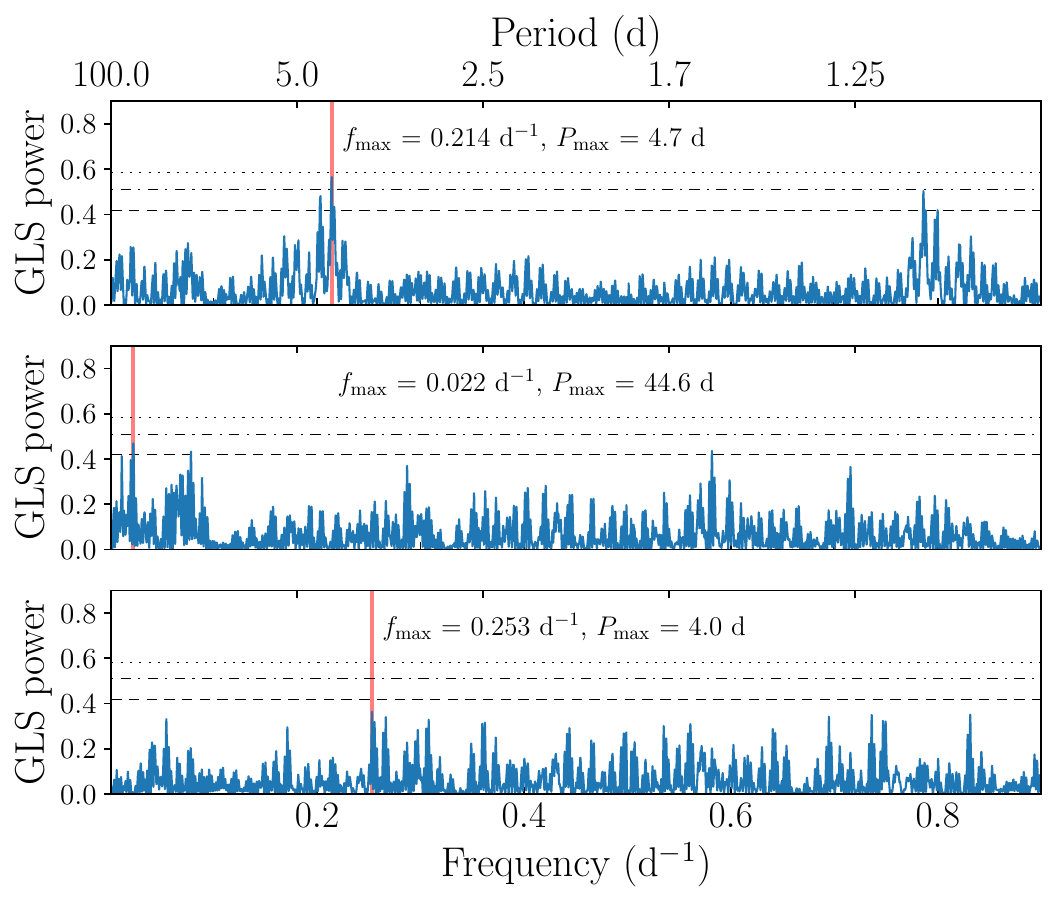}
  \caption{GLS iterative periodogram of ESPRESSO RVs for TOI-912. At each iteration, the sinusoidal model corresponding to the most significant peak is removed. The second signal identified at $\sim 45$~d is related to the rotational period of the star (Sect.~\ref{sec:star_activity}).
  }
    \label{fig:periodogram_iterative}
\end{figure}

We report in Fig.~\ref{fig:asas_sn_toi521} and \ref{fig:asas_sn_TOI912} the GLS periodograms of the photometric data  analysis for TOI-521 and TOI-912, respectively. Figure \ref{fig:TOI521_IRD_HARPS_periodogram} shows the GLS periodogram of the RVs and activity indicators for the IRD dataset (TOI-521) and the HARPS dataset (TOI-912). The stellar rotational period is not identified in any of them. Figure~\ref{fig:periodogram_iterative} shows the iterative periodogram of the ESPRESSO RVs for TOI-912.

\section{RV analysis}\label{sec:app:rv_analysis}
\begin{figure}
\centering
  \includegraphics[width=\linewidth]{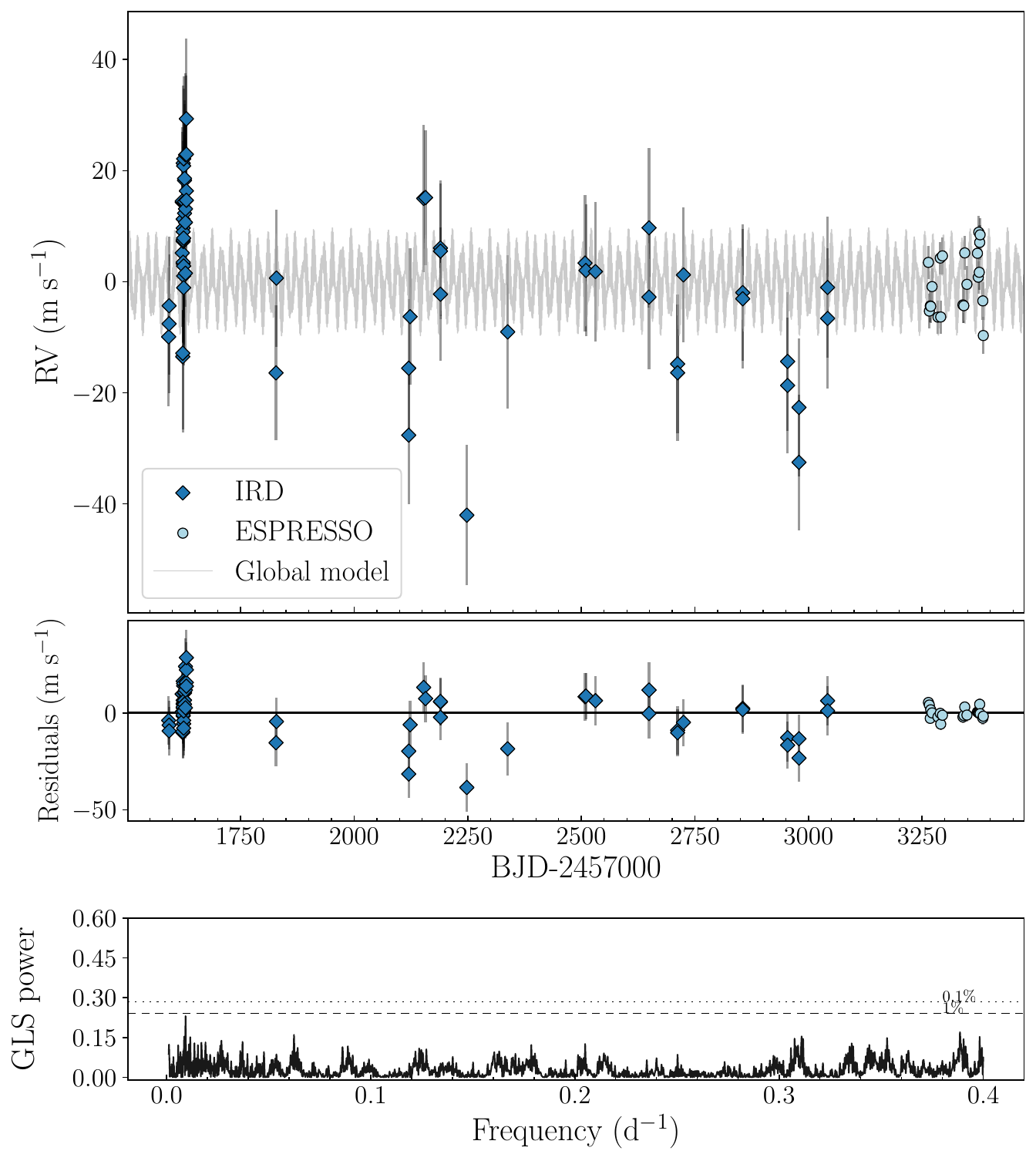}
  \includegraphics[width=\linewidth]{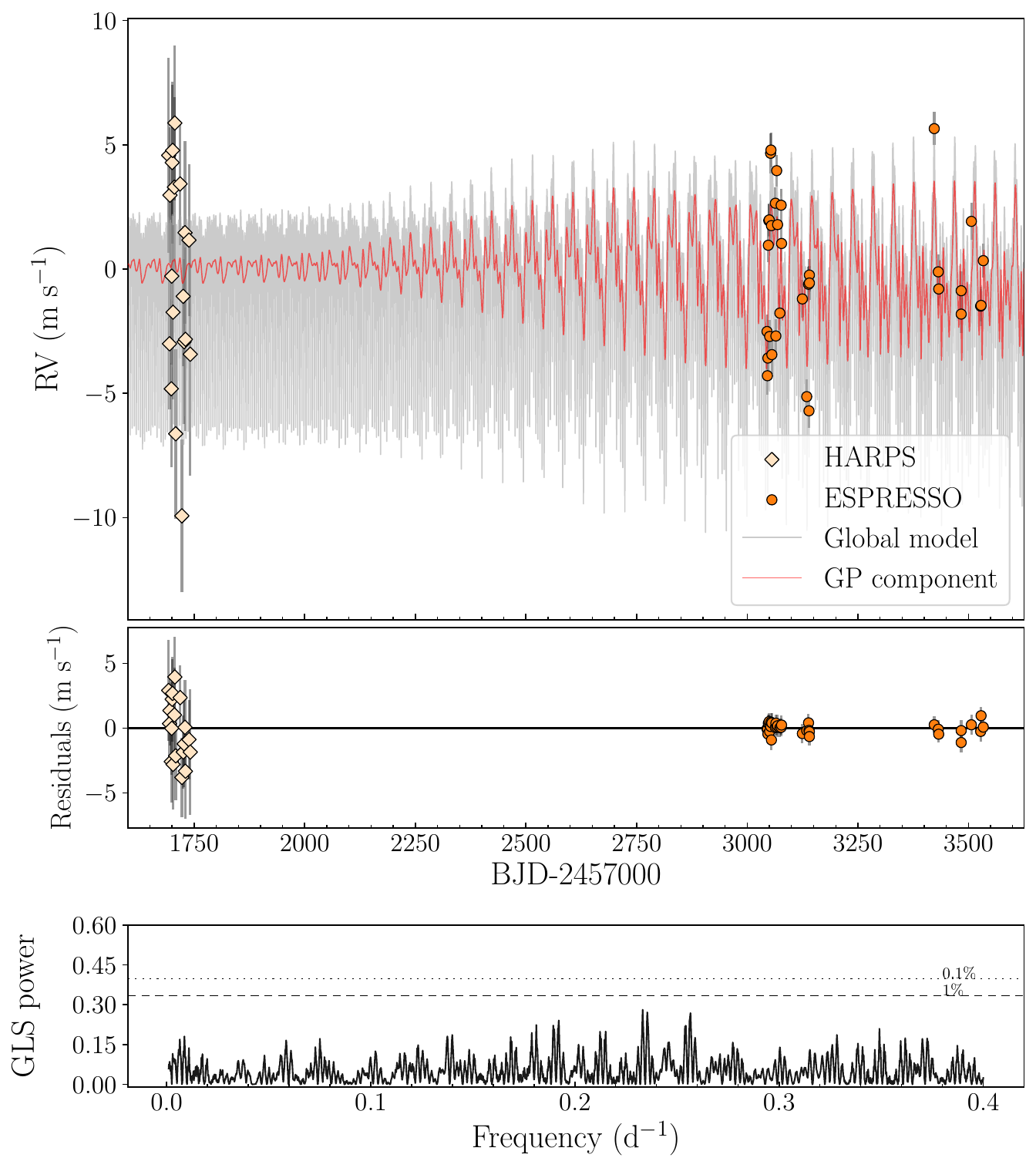}
  \caption{TOI-521 (top) and TOI-912 (bottom) RV model with residuals from the joint fit. The global model and the GP component are shown with solid grey and red lines, respectively. For each time series, the bottom panel shows the GLS periodogram of the residuals, with the 0.1\% and 1\% FAP levels shown as horizontal dotted and dashed lines, respectively. 
  }
    \label{fig:global_rv}
\end{figure}

We report here the results of the global RV modelling of the two stars (Fig.~\ref{fig:global_rv}) as performed in Sect.~\ref{sec:joint_fit}. 

\section{TTV search}
\begin{figure}
\centering
  \includegraphics[width=\linewidth]{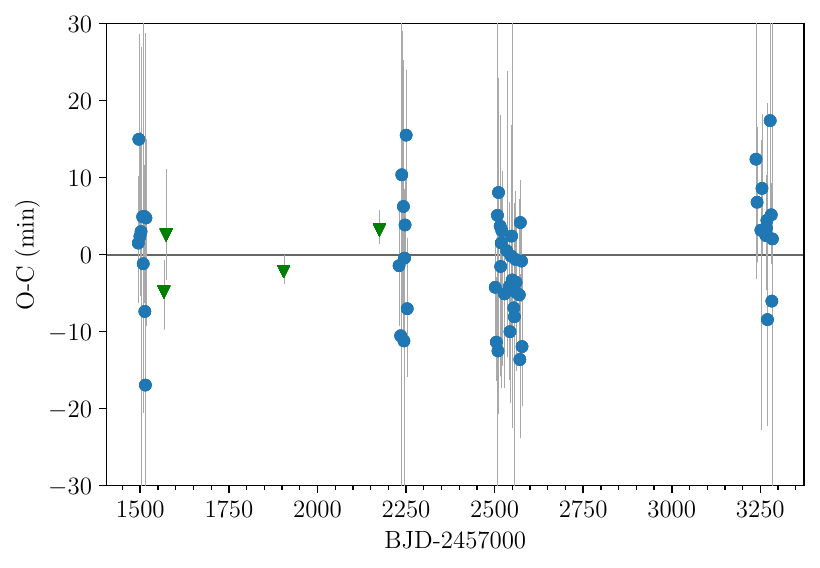}
  \includegraphics[width=\linewidth]{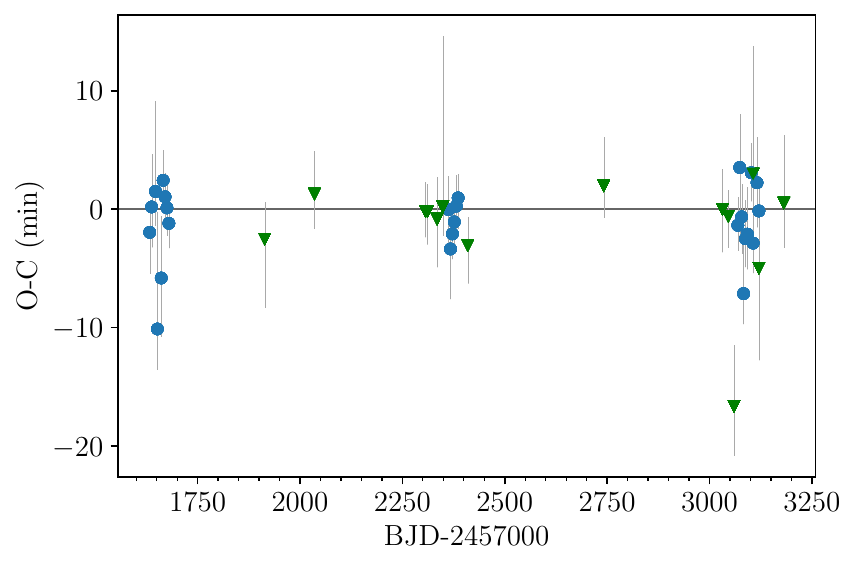}
  \caption{Observed ($O$) minus calculated ($C$) transit times of TOI-521 b. Blue points show the \textit{TESS} transits, and green triangles mark the ground-based photometric data. For TOI-521 b, due to the shallow depth of the transit, the uncertainties on the $T_0$ of individual \textit{TESS} transits are considerable. 
  }
    \label{fig:ttv}
\end{figure}

We show in Fig.~\ref{fig:ttv} the results of the TTV analysis of the transiting planets TOI-521 b and TOI-912 b as described in Sect.~\ref{sec:joint_fit}.

\section{Internal composition}
\begin{figure*}
\centering
\includegraphics[width=0.4\textwidth]{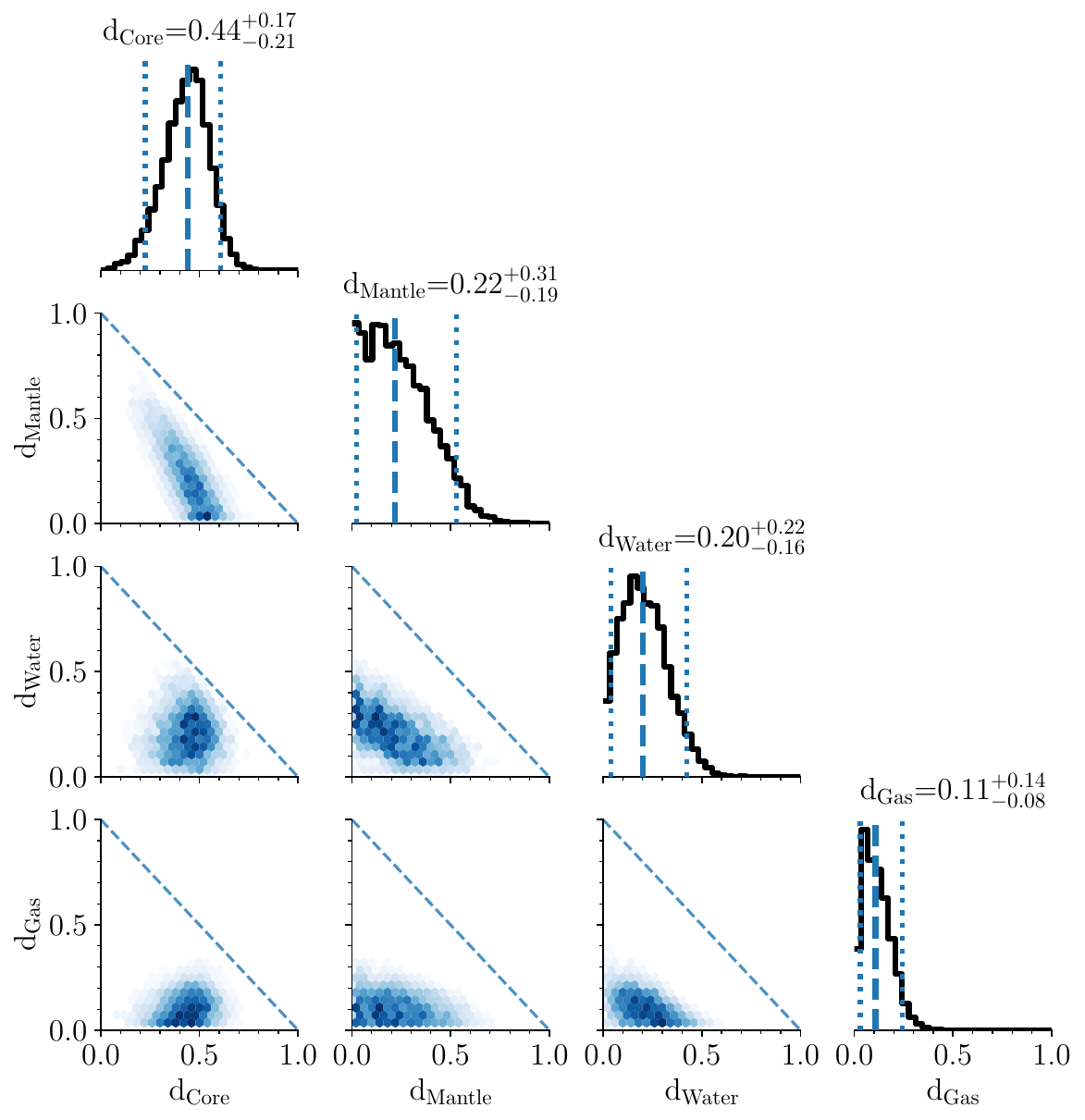}
\includegraphics[width=0.4\textwidth]{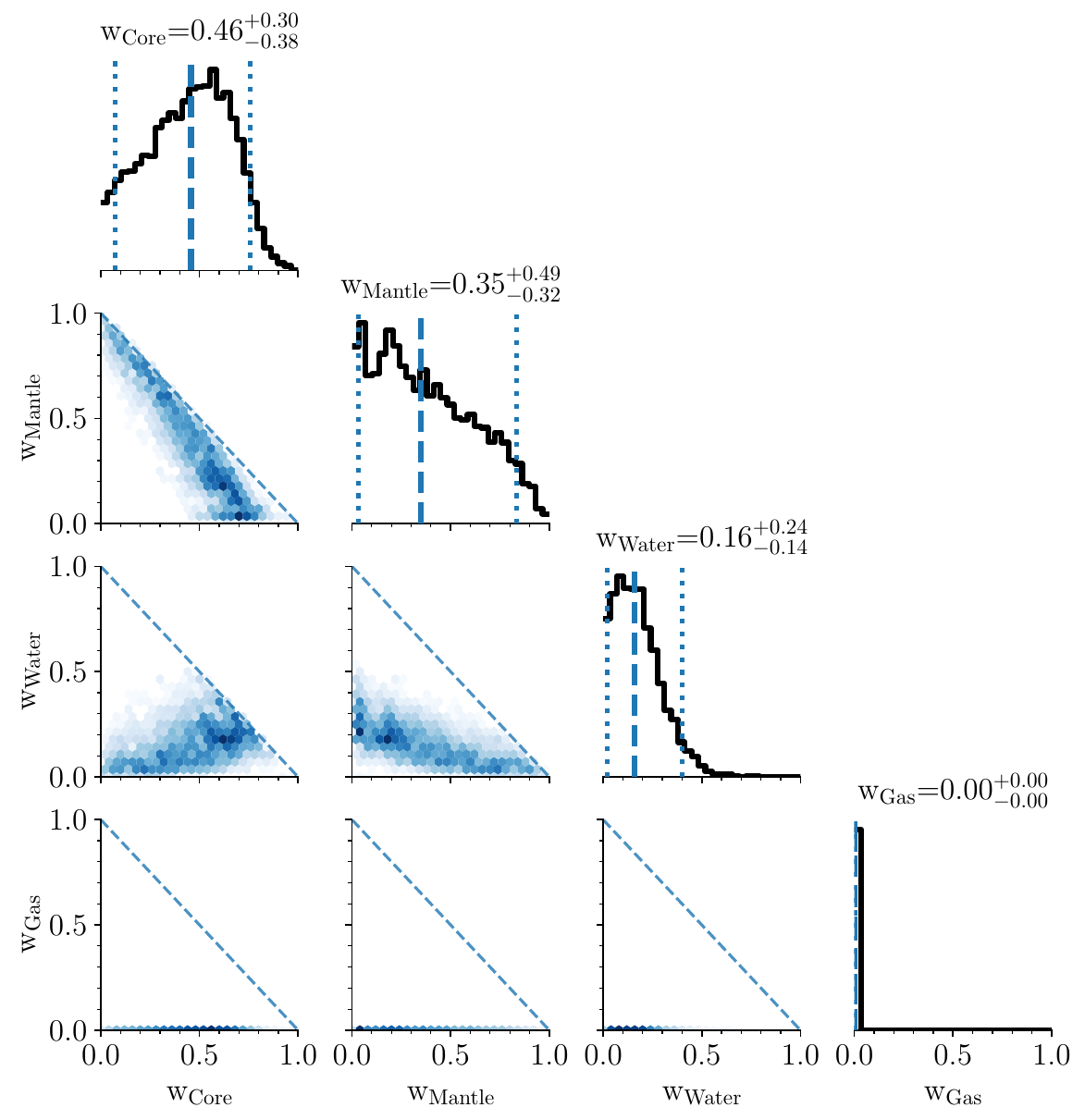}
\includegraphics[width=0.4\textwidth]{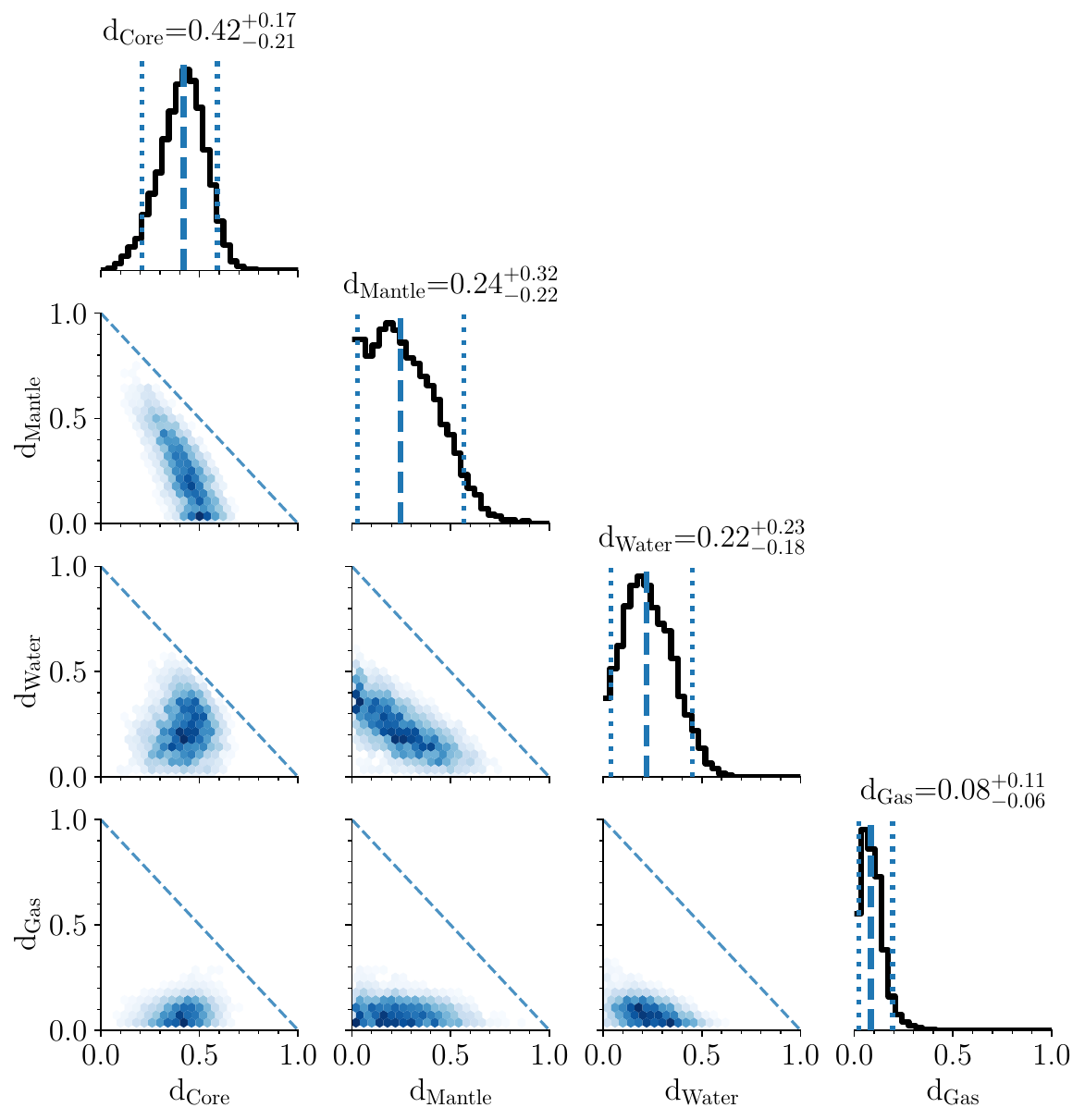}
\includegraphics[width=0.4\textwidth]{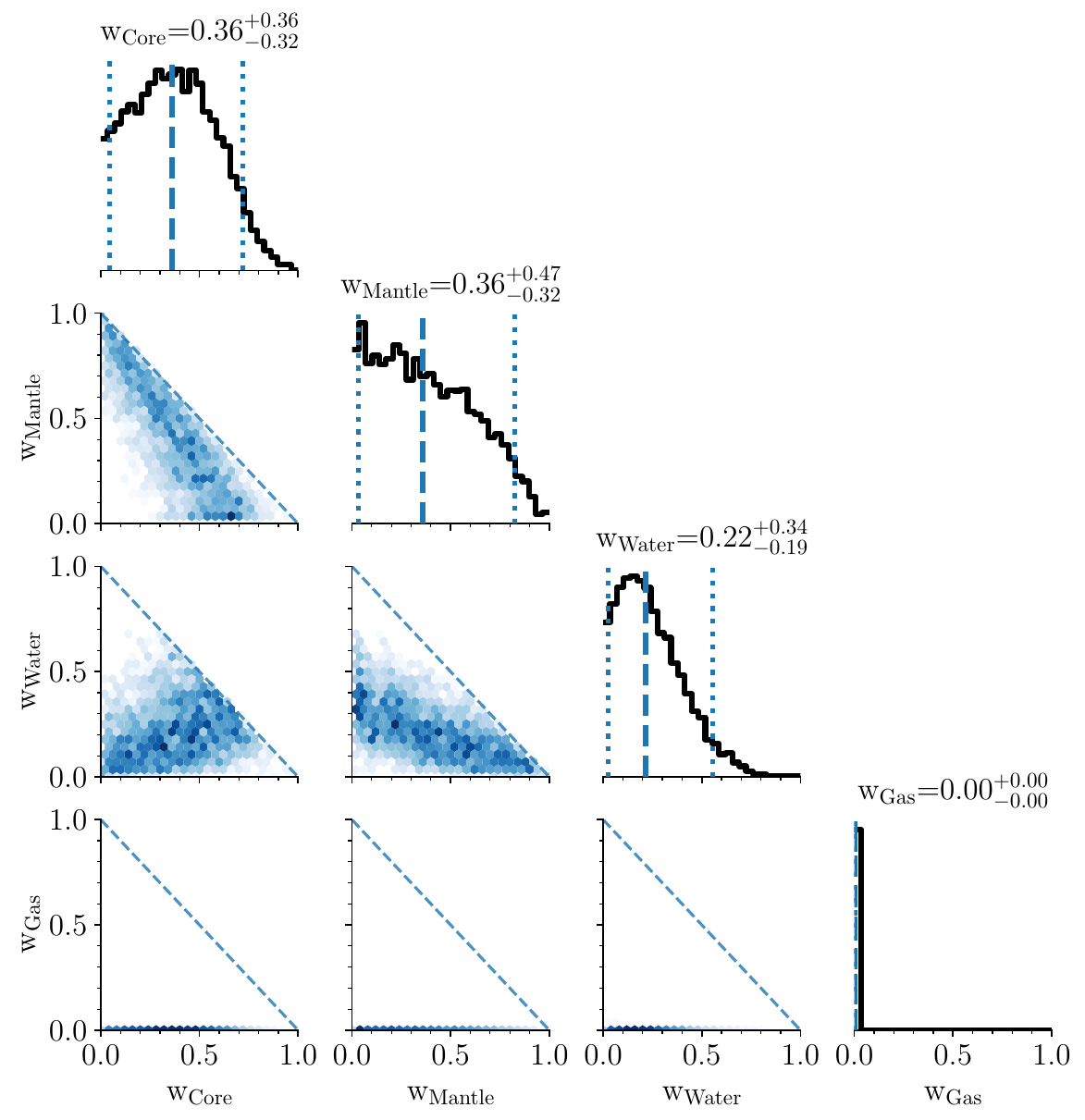}
\caption{Radius fraction (left) and mass fraction (right) of the four interior layers of TOI-521 b (top row) and TOI-912 b (bottom row) computed with \texttt{ExoMDN}.}\label{fig:ExoMDN}
\end{figure*}

We show in Fig.~\ref{fig:ExoMDN} the results of the internal modelling of the planets under analysis performed with \texttt{ExoMDN}.

\end{appendix}

\end{document}